\documentclass[referee]{aastex}

\newcommand{\kms}{\mbox{km\,s$^{-1}$}}

\newcommand{\Msun}{\mbox{M$_{\odot}$}}

\def\HII        {\hbox{H \small{II}}}

\usepackage{graphicx}
\usepackage{txfonts}
\usepackage{natbib}

\begin{document}
\title{Ion and neutral molecules in the W43-MM1(G30.79 FIR 10) infalling clump}

\author{Paulo C.\,Cortes\altaffilmark{1}}
\affil{ National Radio Astronomy Observatory - Joint ALMA Office, Alonso de Cordova 3107, Vitacura, Santiago, Chile}
\email{pcortes@alma.cl}

\begin{abstract}
The high mass star forming clump W43-MM1 has been mapped in N$_{2}$H$^{+}(4 \rightarrow 3)$,
C$^{18}$O$(3 \rightarrow 2)$, SiO$(8 \rightarrow 7)$, and in a single pointing in DCO$^{+}(5 \rightarrow 4)$
towards the center of the clump. Column densities from these observations
as well as previous HCO$^{+}(4 \rightarrow 3)$, H$^{13}$CO$^{+}(4 \rightarrow 3)$,
HCN$(4 \rightarrow 3)$, H$^{13}$CN$(4 \rightarrow 3)$, and CS$(7 \rightarrow 6)$ data, 
have been derived using the RADEX code, results later used to derived 
chemical abundances at selected points in the MM1 main axis. 
Comparing with chemical models, we estimate an evolutionary age of $10^{4}$ years for
a remarkable warm hot core inside MM1. We also proposed that the dust temperature 
derived from SED fitting in MM1 is not representative of the gas temperature deep
inside the clump as dust emission may have become optically thick.
By deriving a deuterium fractionation of $1.2 \times 10^{3}$, we estimate an electron fraction
of $X(e)= 6.5 \times 10^{-8}$.  Thus,
the coupling between the neutral gas and the magnetic field is estimated by computing the ambipolar 
diffusion Reynolds number $R_{m}=18$ and the wave coupling number $W=110$.
Considering that the infalling speed is slightly supersonic (M=1.1) but
sub-alfvenic, we conclude that the MM1 clump has recently or is in the process of decoupling the
field from the neutral fluid. 
Thus, the MM1 clump appears
to be in an intermediate stage of evolution in which a hot core has developed
while the envelope is still infalling and not fully decoupled from the ambient magnetic field.
\end{abstract}


\maketitle

\section{Introduction}\label{se:INTRO}

High mass star formation involves the study of gas masses many orders of magnitude 
larger than low mass stars increasing the complexity as
the process is highly energetic and dynamic. The distances at which the
regions of interest are located  are usually
kilo-parsecs away from the sun, complicating the use of single dish telescopes
as current dish sizes do not produce
enough angular resolutions to resolve individual cores. 
Thus, the emission is often diluted within the beam complicating the
interpretation as the high mass cores inside clumps are usually shrouded by
optically thick envelopes, even with dust emission. 
We know that massive stars 
($\gtrsim$10\,\Msun) form in giant molecular clouds several orders
of magnitude larger than the low-mass star forming clouds. As it is still
not clear how these giant ensembles of gas came to be, we know that 
their location are correlated with the spiral arms of our galaxy. These
massive complexes are known to harbor cores with usually several sources
as massive stars are born in groups or associations. Once a massive star
turns on it will quickly perturb and
ionize the surrounding envelope creating what it is
known as a hot core \citep{Wood1989,Hoare2007}.  
The intense radiation coming from these young sources, 
will evaporate C and O bearing molecules
from ice mantles surrounding dust grains, changing dramatically the
chemical diversity around the star(s) \citep{Herbst2009}.
Even though,
the inventory of complex molecules in hot cores have been growing
consistently in the past 10 years, still chemical and dynamical 
models have problems trying to reproduce abundances and 
explain the chemical differentiation seen in high mass cores \citep{Garrod2008}.
Thus, probing the densest regions  becomes paramount to understand
their physical and chemical conditions. On the otherhand, the 
complete set of physical parameters have to be considered if we want to
fully understand the process. Magnetic fields are known to be ubiquitous
in the ISM and giant molecular clouds are no exception. Strong magnetic field
strengths have been measured from CN Zeeman splitting towards high mass star forming regions
to be in the order of milligauss \citep{Crutcher1996,Crutcher1999a}.
Also, observations done from polarized dust emission find ordered
field structure with strength estimations around the same values \citep{Lai2001,Cortes2005,Cortes2006a,Cortes2008}.
Thus, incorporating the magnetic field along with the set of parameters
used to study the star formation process is critical.

\noindent In this paper we present new 
observations of DCO$^{+}(5 \rightarrow 4)$, N$_{2}$H$^{+}(4 \rightarrow 3)$, and 
SiO$(8 \rightarrow 7)$ as well as re-processed\footnote{By re-processing, we mean using the
data cubes obtained by PCI to produce maps and calculate column densities with RADEX (see 
section \ref{abundances} which have not been done previously} HCO$^{+}(4 \rightarrow 3)$,
H$^{13}$CO$^{+}(4 \rightarrow 3)$, HCN$(4 \rightarrow 3)$, H$^{13}$CN$(4 \rightarrow 3)$,
and CS$(7 \rightarrow 6)$ observations done previously by \citet{Cortes2010} and
hereafter PCI, from the high mass star forming clump W43-MM1 (also known as G30.79 FIR 10).
The W43-MM1 is a large high mass star forming clump located within the W43 region  
near $l=31^{\circ}, b=0^{\circ}$.
\citet{Liszt1995} observed the W43 molecular complex in HCO$^{+}$ and $^{13}$CO
finding a series of several rings and shells in the dense  molecular gas, which
they attributed to be a product of star formation.
\citet{Mooney1995} made the first dust continuum observations of this source at 1.3 mm
using the IRAM 30-meter telescope. They detected a total flux of 13.6 Jy positioning
MM1 at an interface with the extended \HII\ region in the W43 complex.
Several H$_{2}$O masers \citep{Cesaroni1988} have also been detected within an 
arcsecond from the dust peak found in the 1.3 mm data. 
No centimeter radio-continuum emission seems to be associated with
MM1, which suggests that the source is in an early stage of evolution.
Additional continuum mapping has been done by \citet{Motte2003} 
at 1.3 mm and 350 $\mu m$ with the IRAM 30-m and CSO telescopes\footnote{IRAM stands for 
Institut de Radioastronomie Millimetrique, and CSO stands for Caltech Sub-millimeter Observatory}
respectively.
They determined a v$_{lsr}= 98.8$ km s$^{-1}$ from H$^{13}$CO$^{+}$ and
 a dust temperature of T$_{\textnormal{dust}} \sim 19$ K, by fitting 
a grey body model to the spectral energy density distribution. \citet{Cortes2006a}
made interferometric observations of polarized dust emission with BIMA,
finding an ordered pattern for the field and estimating and deriving 
a total mass of about 3000 \Msun for the MM1 clump. Finally, PCI studied the
gravitational equilibrium of this source finding compelling evidence for
infalling motions and refining a previous estimation of the magnetic field
strength in the plane of the sky \citep{Cortes2006a} to $B = 855$ $\mu$G.
PCI also produced several maps from the previously mentioned molecular species.
From this set of observations, we have derived chemical abundances as well as
spatial distributions of molecules, which we use to estimate the electron
fraction and the coupling parameters between the magnetic and the neutral gas. 
The paper is organized as follows, Sect. 1 is the
introduction, Sect. 2 presents the observation configuration, Sect. 3 the
observational results, while in
Sect. 4 we discuss the chemical abundances, the spatial
differentiation, the fractional ionization, and the ion-neutral coupling. 
Finally, Sect. 5
presents the summary and conclusions.

\section{Observation Configuration}\label{OBS}

\subsection{ASTE Observations}\label{se:ASTEOBS}
The W43-MM1 clump was observed during July 2010 using the Atacama 
Sub-millimeter Telescope
Experiment (ASTE) from the National
Astronomical Observatory of Japan (NAOJ)
\citep{Kohno2005}.  The telescope is located at {\em Pampa la bola} in
the Chilean Andes plateau reserve for Astronomical research at 4900
meters of altitude.  ASTE is a 10\,m diameter antenna equipped with a
345\,GHz double side band SIS-mixer receiver. We
observed DCO$^{+}(5 \rightarrow 4)$ and N$_{2}$H$^{+}(4 \rightarrow 3)$ 
by tuning the CAT345 receiver to the appropriate frequencies (see Table \ref{tab1} for
frequencies and map sizes)
giving a beam size of $\sim 20^{\prime \prime}$, and a velocity resolution
of 0.1\,\kms. The MAC (a XF-type digital spectro-correlator) was set to high resolution mode with
a total bandwidth of 128 MHz.  The pointing accuracy was
in the order of 2$^{\prime \prime}$, with VY-CMa used as the pointing
source. The observations were done by performing single pointing (position
switching) with an off position carefully selected to avoid contamination
to the on position. Also, a small $60^{\prime \prime} \times 60^{\prime \prime}$ 
map for N$_{2}$H$^{+}$ was obtained centered at the reference position.
The observations were done under excellent weather conditions  
(precipitable water vapor or $<$PWV$>$ $~$ 0.5\,mm and wind speed under 3 m/s). 
Our reference position was
$(\alpha,\delta)=(18^{\mathrm{h}}47^{\mathrm{m}}46.9^{\mathrm{s}},
-1^{\circ}54^{\prime}29.1^{\prime \prime})$ (J2000), which coincides
with the peak dust emission reported and used by
\citet{Mooney1995,Motte2003,Cortes2006a,Cortes2010}; while the OFF position used is
$(\alpha,\delta)=(18^{\mathrm{h}}49^{\mathrm{m}}46.7^{\mathrm{s}},
-2^{\circ}00^{\prime}52.0^{\prime \prime})$ 
Initial data reduction and
calibration was done using the NEWSTAR package, and the calibrated data were
later exported into our own software for analysis and plotting.

\subsection{APEX observations}\label{APEXOBS}
Observations were performed during the first week of August 2008
using the Swedish Heterodyne Facility Instrument (SHFI) mounted on the
Atacama Pathfinder Experiment telescope (APEX) \citep{Gusten2006b}, located
at llano de Chajnantor also in the Chilean Andes plateau reserve for Astronomical research at. 
The SHFI was tuned to 347.33 GHz to detect the SiO$(J=8 \rightarrow 7)$ molecular transition.
The spectrometer was set up
to 8192 channels with a resolution of 0.1\,km\,s$^{-1}$.  
beam efficiency is $\eta$=0.73$\pm$0.07 as measured by the APEX staff,
with a pointing accuracy better than 2$^{\prime \prime}$ and a
beam size of $19^{\prime \prime}$.  The APEX wobbler was not used
for this observations.
Jupiter and R-Aql were used as intensity and
pointing calibrators where the observations were calibrated by the usual
chopper-wheel method. The observations were done in raster mode with
spacings of $15^{\prime \prime}$ from the same reference position used
for the ASTE observations. The initial data reduction was done with
the GILDAS-CLASS reduction package and the final analysis with our own
software tools.

\begin{deluxetable}{c c c c c c c c c c}        
\tabletypesize{\scriptsize}
\rotate
\tablecaption{Observing parameters for molecular lines observations towards W43-MM1.\label{tab1}}        
\tablehead{
\colhead{Line} & \colhead{Transition} & \colhead{Frequency} & \colhead{Beam Size} & 
\colhead{Area mapped} & \colhead{Peak} & \colhead{Line-width} & \colhead{$\int$T$dv$} & 
\colhead{$\sigma$} & \colhead{Telescope} \\
\colhead{ } & \colhead{ } & \colhead{[GHz]} & \colhead{[arcsec]} & \colhead{ } & 
\colhead{[K]} & \colhead{[\kms]} & \colhead{[Kkm/s]} & \colhead{[K]} & \colhead{ }
}
\startdata  
   DCO$^{+}$        & $(J=5\rightarrow4)$ & 360.1698810 & 20.9 & single pointing & 0.044 & 2.9 & 0.11 & 0.01 & ASTE\\ 
   N$_{2}$H$^{+}$   & $(J=4\rightarrow3)$ & 372.6725090 & 20.3 & $60^{\prime \prime} \times 60^{\prime \prime}$ & 2.1 & 3.7 &  7.3 & 0.08 & ASTE \\
   SiO              & $(J=8\rightarrow7)$ & 347.3305786 & 21.7 & $60^{\prime \prime} \times 60^{\prime \prime}$ & 0.6 & n/a &  n/a & 0.24 & APEX  \\
\enddata
\end{deluxetable}

\section{Results}
\subsection{Neutrals Observations}\label{se:C18O}

\noindent In PCI, we reported a single $30^{\prime \prime} 
\times 30^{\prime \prime}$  averaged spectrum, 
centered at the MM1 reference position, for C$^{18}$O$(3 \rightarrow 2)$. Even though we mapped
a larger area of $150^{\prime \prime} \times 120^{\prime \prime}$ sampled every $15^{\prime \prime}$,
 we did not present the complete set of observations in that work. Here, we 
present the integrated intensity map in Figure \ref{c18omap}.
The map follows the elongated 
structure seen from the $^{12}$CO$(3 \rightarrow 2)$ map presented in PCI (see Figure 2 there).
Two clear peaks are seen in the C$^{18}$O map, one is coincident 
with the reference position and the other is located at
$(\alpha,\delta)$=($-120^{\prime \prime},-45^{\prime \prime}$).  The first peak
agrees with the peak in the dust emission, HCN$(4 \rightarrow 3)$,
HCO$^{+}(4 \rightarrow 3)$, and CS$(7 \rightarrow 6)$ presented in PCI.
The second and strongest peak is interesting as it is located at the edge of the map following the
$^{12}$CO$(3 \rightarrow 2)$ peak where it
appears to be anti-correlated with the dust emission 
i.e. the dust emission is minimum where CO is maximum. According to the continuum maps 
of MM1 done by \citet{Motte2003} 1.3 mm and 350 $\mu$m, the dust emission 
decreases an order of magnitude with respect to the dust-peak at this position.
As previously suggested by PCI, this area seems to be at an interface
with the \HII\ region produced by a large cluster of massive stars located at
the center of W43 \citep{Wilson1970,Smith1978}. 
Thus, it it possible that we are seeing part of a PDR at that area.

\noindent We observed an area of $60^{\prime \prime} \times 60^{\prime \prime}$
in SiO$(8 \rightarrow 7)$ towards the center of the
MM1 clump (see Figure \ref{siocenter}). 
Large amount of data were discarded as the baselines showed evident
ripples which we could not remove.
There is no conclusive evidence of SiO emission
from these data, only an upper limit of 0.6 K around the
systemic velocity of MM1 is obtained with an rms noise of 0.24 K which
is less than $3\sigma$. Thus, we are not able to confirm the marginal evidence 
find by PCI from the HCO$^{+}$ and CS line-wings.
However, new un-published SMA\footnote{
SMA stands for Sub Millimeter Array} interferometric
observations  \citep{Sridharan2011}
found evidence for outflow emission towards MM1 as the interferometer
filters the envelope probing angular scales closer to the center of the core.
 Thus, it is likely that
the outflow is young and deeply embedded into the clump with its emission diluted
within the APEX beam.

\subsection{Ion observations}\label{se:DCOp}

Single pointing observations were obtained towards the W43-MM1 core 
in order to detect the DCO$^{+}(5 \rightarrow 4)$ molecular transition.
Figure \ref{1} shows the spectrum obtained and its corresponding
Gaussian fit (in red); Table \ref{tab1} summarizes the main 
molecular line parameters.
With a single peak of 44 mK over a noise level of
$\sigma=10$ mK, the line intensity corresponds to a 4$\sigma$ 
significance observation, which is clearly a detection. 
The data were processed by keeping only spectra with stable 
baselines using order 1 polynomial fits. The spectrum was binned
every 10 channels, giving a velocity resolution of 1 \kms.
The DCO$^{+}$ molecule has been detected in its lower transition towards a 
number of star forming cores with similar results to our own \citep{Wooten1979,Caselli2002a,Caselli2002b}.
The line profile is clearly Gaussian, which suggests that the line
is likely optically thin. Even though it is expected to find 
high abundances of DCO$^{+}$ in early stages of massive cores, the abundances 
are unlikely to be large enough to make the line optically thick.
By looking at Figure 1 we also noticed a strong line besides the
DCO$^{+}$(5-4) detection (with a Gaussian fit in blue). 
The corresponding line center frequency from
the best fit is 360.1847 GHz,  it is currently un-identified  and its nature will be discussed
in section \ref{se:uniden}.  



\noindent We observed N$_{2}$H$^{+}(4 \rightarrow 3)$ rotational transition
towards the center of MM1.
A $60^{\prime \prime} \times 60^{\prime \prime}$ integrated emission 
map is presented in Figure \ref{n2hpmap}. The map is
centered at the reference position and the emission was sampled
every $10^{\prime \prime}$ in both $\alpha$ and $\delta$. 
Also, a short 5 minutes
scan was done at the center to increase the signal to
noise in the spectrum. The spectrum, shown in Figure \ref{n2hpcenter}
with its corresponding Gaussian fit, was binned every 10 channels
giving a velocity resolution of 1 \kms.
From the fit, we obtained a peak intensity of 2.1 K with a rms noise
of $\sigma=80$ mK. No hyperfine structure is seen as is likely blended due to
a large line-width of $\Delta v=7.5$ \kms seen in the emission, which has also been seen
by others (\citealt[][for  N$_{2}$H$^{+}(4 \rightarrow 3)$]{Friesen2010}  
and \citealt[][for N$_{2}$H$^{+}(3 \rightarrow 2)$]{Fontani2006}). 
The N$_{2}$H$^{+}$ center spectrum 
seems to be consistent with an optically thick line centered at 
$v=$97.14 \kms, which is blue-shifted from the core v$_{lsr}$ and 
in agreement  with infalling evidence presented in PCI and by
others for similar high mass star forming regions \citep{Fuller2005}. 
The N$_{2}$H$^{+}$ integrated intensity map
appears as a filament from South-East to North-West, which is consistent
with the $^{12}$CO$(3 \rightarrow 2)$,
HCN$(4 \rightarrow 3)$, and HCO$^{+}(4 \rightarrow 3)$ mapping done in PCI (see Figures 2, 5, and 8 in
that work) and our C$^{18}$O$(3 \rightarrow 2)$ map.
This filament is also suggested by the continuum emission maps at 
1 mm and 350 $\mu m$, and integrated intensity HCO$^{+}(3 \rightarrow 2)$
maps of \citet{Motte2003}. A common correlation found in all these maps is
that the peak integrated emission is coincident with the dust peak, where the
dust peak is located at the reference position used here and taken from our 
previous interferometric observations on this source
\citep{Cortes2006a}. However, in our N$_{2}$H$^{+}$ map  
the peak is offset one beam from the reference position. 
At this frequency, the ASTE beam is around $20^{\prime \prime}$
which coincides with the core size inferred from the interferometric data.
This suggests that the emission from this point in the
map come from independent regions of the MM1 clump, which at the assumed
distance of 5.5 kpc corresponding to 0.5 pc. 
Comparing with the HCO$^{+}$ map, we see that the emission is seen
enclosing a larger area around the dust peak, which includes the N$_{2}$H$^{+}$
peak ($T\Delta v$ is over 45 K \kms at the innermost contour). The integrated
intensity quickly decreases north of the peak in both maps.
In our previous maps of
HCN$(4 \rightarrow 3)$ and CS$(7 \rightarrow 6)$ we saw  little or almost  no emission
offset from the center suggesting that most of the dense neutral gas was unresolved and
inside the MM1 core. Even though CO emission is expected to be widespread,
a similar situation is suggested in the C$^{18}$O map where the
dust peak coincides with a local C$^{18}$O peak at the same position, quickly decreasing
outside the dust peak. Figure \ref{n2hppanel}
shows the N$_{2}$H$^{+}(4 \rightarrow 3)$ spectra towards selected offsets from the reference
position corresponding to the brightest points in the map. All three spectra have higher
intensities than the spectra taken at the center of the map. Similar situations
have been in seen in low and high mass star forming regions where N$_{2}$H$^{+}$
appears to be depleted towards the dust peaks 
\citep{Daniel2007,Pirogov2007,Zinchenko2009,Miettinen2009,Johnstone2010,Busquet2011}.
It is likely that chemical differentiation has depleated 
 N$_{2}$H$^{+}$ at the center of MM1.

\section{Discussion}\label{se:disc}

\subsection{The un-identified line in the DCO$^{+}$ spectrum}\label{se:uniden}

To explore the nature of the un-identified line seen
besides the DCO$^{+}(5 \rightarrow 4)$ spectrum, we first considered the 
possibility of a spurious instrumental signal. The ASTE telescope
has a cartridge-type side-band separating the (2SB) mixer receiver 
working at the 350 GHz band. The image rejection ratio  is 15 dB,
which makes opposite sideband leakage unlikely. The telescope
receiver has been carefully tested and no spurious signals are
reported at that frequency \citetext{T. Sakai, priv.\ comm.}. 
The DCO$^{+}$ line was observed in
the upper sideband of the ASTE receiver. Thus, we searched for
strong lines in the lower sideband (12 GHz away), which may have
leaked into the upper sideband. 
Leaking from the lower
sideband would require a strong line due to the high side-band image rejection ratio
and according to the splatalogue 
database for molecular spectroscopy, no strong lines are present
in a 500 MHz interval centered at 348.169 GHz. 
 Also, contamination from
atmospheric lines is also ruled out as the position switching 
removes any atmospheric contribution. Another possibility is
that the line is the result of a baseline ripple. These instabilities
are usually the result of a standing wave leaked, due to reflections
between the feed and the antenna structures, into the receiver.
A possibility is oscillations of the goretex membrane that
covers the receiver cabin under the dish. To explore
this possibility, the 484 scans gathered were inspected one by one. Those with
clear ripples were flagged-out and not used. After baseline
removal, the remaining scans were inspected again looking for
further instabilities. Only those with flat baselines were kept, which
imply removing about 30\% of the observations. However even after
removing spectra with ripples, it cannot be ruled out that the line feature
is the result of such instability. Having established this, 
we consider the possibility of the signal being real. If the feature is
a molecular transition, its rest frequency would be 360.1847 GHz.
The closest molecular transition is the $^{13}$C isotopologue of formic acid,
in the $16( 3,14) \rightarrow 15( 3,13)$ at 360.18795 GHz, as reported by the
splatalogue database but originally provided by CDMS \citep{Muller2005}. However,
the emission seen here is too strong even for $^{12}$C formic acid \citep{Arce2008}
and thus, this possibility is unlikely.
On the other hand, there is always the 
possibility that we are seeing a transition from an un-identified molecule.
Another scenario is that this line corresponds to DCO$^{+}$ emission
at a different velocity. If that is the case, it will mean strong DCO$^{+}$
emission from a cold component in the line of sight. This would be an exciting new result as 
deuterium chemistry is thought to be enhanced in cold cores by 
freezing out of carbon bearing molecules on grain surfaces \citep{Bergin2007}.
Nevertheless, we cannot conclude with certainty
what the nature of the emission seen here is. It is difficult to correlate with emission
from other molecules as their line-width is large, specially HCN.
Only more independent observations will help confirming this un-identified emission.

\subsection{Column densities and abundances}

Assuming that the clump is in general homogeneous and isothermal, it is
possible to obtain information about the optical depths and column 
densities by analyzing the line intensities and the velocity dispersions.
Based on these assumptions, we used the radiative transfer code RADEX \citep{Schoier2005}
to estimate the column densities and abundances for several molecules including some of the ones
presented in PCI (see Table \ref{abundances}). The detailed
source geometry is not known as the core is unresolved and the number of sources
inside MM1 is yet to be determined. Thus, we make no assumptions about the 
clump structure or possible number of sources inside. 
As the RADEX code does not assume any kind of source geometry or velocity fields, 
the column density for the emitting material is 
adjusted until the line intensity is matched. We also corrected for beam dilution by multiplying the
output from RADEX by the beam filling factor calculated for each transition where
the source size was obtained from the interferometric observations of \citet{Cortes2006a}.
Because the beam sizes for all molecules are comparable (differences are $\sim$
$2^{\prime \prime}$, see Table \ref{tab1}), the 
correction is small. Thus we should not expect strong beam-diluted emission if
the molecular gas follows the dust distribution. However, there is evidence
for a remarkably warm hot-core inside the MM1 clump with T $\sim$ 400 K and 0.03 pc in size \citep{Sridharan2011}.
As the MM1 hot core is very small in size, any emission
coming only from the hot core will be extremely diluted within any current single dish beam.

We derived column densities for all observed molecules, with the exception of
 HCO$^{+}$ and HCN as the lines are self-absorbed and more
accurate geometric source modeling for the radiative transfer problem should be done.
Instead, we used RADEX for the less abundant H$^{13}$CO$^{+}$ and H$^{13}$CN
to estimate their column densities and then used a $^{12}$C/$^{13}$C 
ratio of 50 as observed in the molecular ring \citep{Wilson1994} to derive the
column densities for the main isotopomers.
The N$_{2}$H$^{+}$ column density was derived using the collision rates
provided by RADEX where hyperfine structure is considered in the calculation.
However, the hyperfine structure appears to be blended in the main emission as the line-width
is larger than the hyperfine components separation. The remaining molecular column
densities where calculated using RADEX and then were compared with
classical column density estimations \citep{Caselli2002b}, agreement was within a factor of 2.

\begin{figure}
\centering
\includegraphics[width=0.9\hsize]{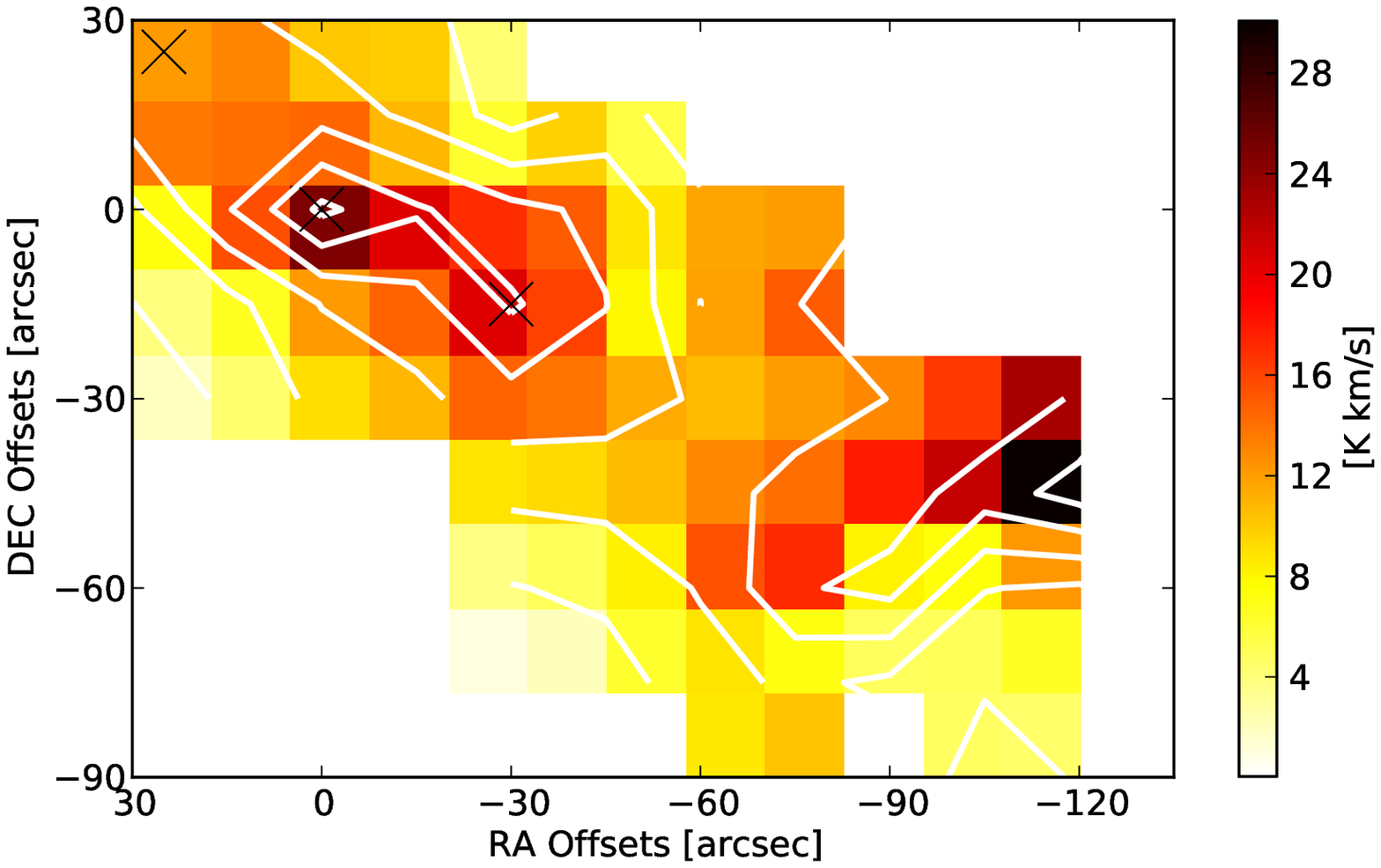}\caption{
A $60^{\prime \prime} \times 60^{\prime \prime}$ integrated intensity map
of C$^{18}$O$(3 \rightarrow 2)$ from W43-MM1 centered at the dust peak. The color scale
correspond to the integrated emission after applying a {\em nearest} interpolation
scheme. The contours are choosen to be 8, 12, 16, 20, and 24 K \kms 
while the colorscale is indicated in the colorbar. Crosses indicate the positions at which we
derived fractional abundances from our sample of molecules. The offsets are $(30^{\prime \prime}, 
30^{\prime \prime})$, $(0^{\prime \prime}, 0^{\prime \prime})$, and $(-30^{\prime \prime}, 15^{\prime \prime})$ in $(\alpha,\delta)$. 
Note tha the reference position is offset from the center
of the map by  $60^{\prime \prime}$ in $\alpha$ and $30^{\prime \prime}$
in $\delta$.
}
\label{c18omap}
\end{figure}

\begin{figure}
\centering
\includegraphics[width=0.9\hsize]{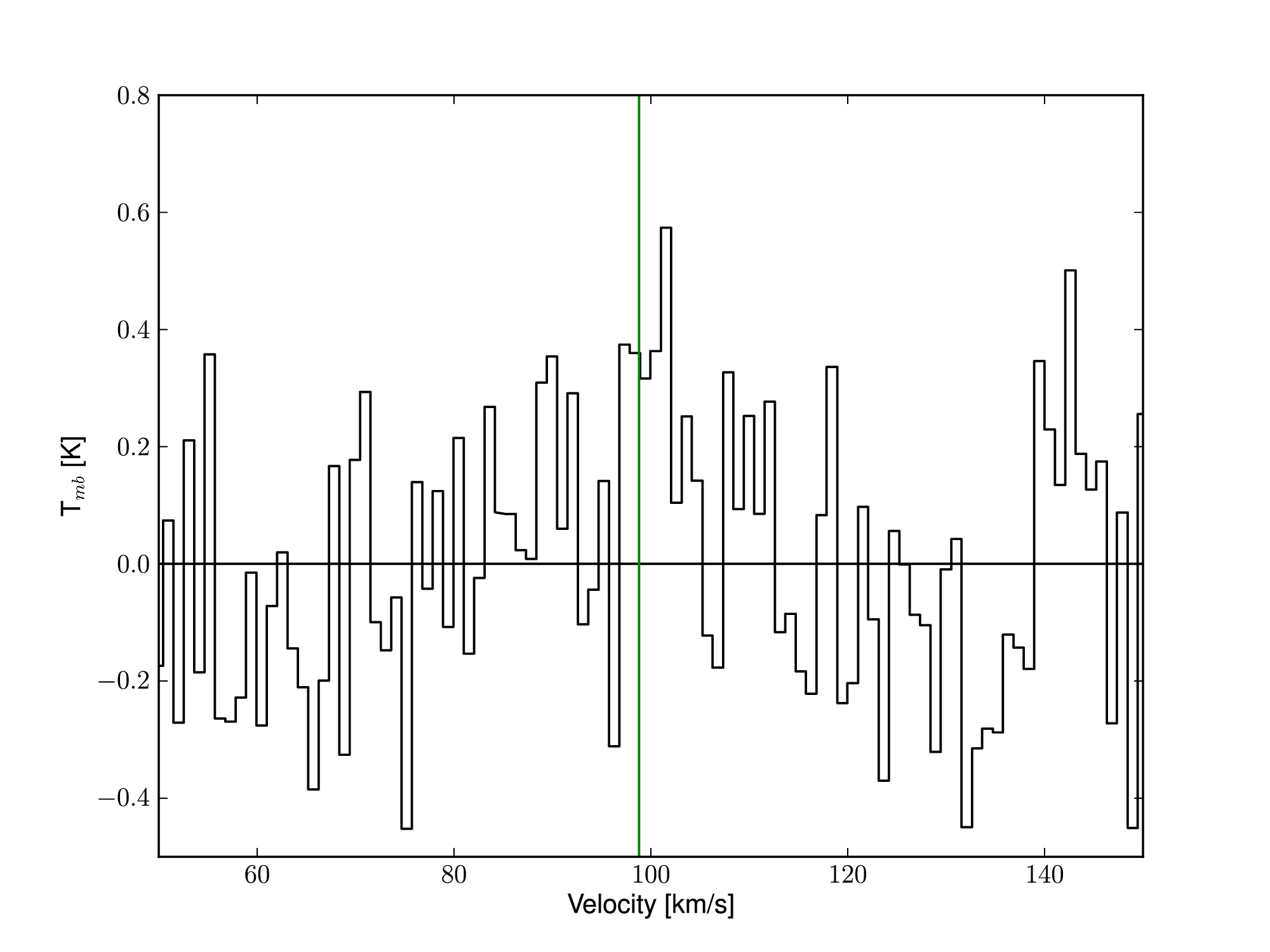}\caption{
The figure shows our SiO$(8 \rightarrow 7)$ spectrum towards W43-MM1.
The black line marks the zero baseline and  the systemic
velocity of the clump, 98.9 \kms is shown by the green line. 
The spectra is binned every 10 channels giving a resolution in
velocity of about 1 \kms giving and rms noise of 0.24 K. The intensity upper limit
at the systemic velocity is 0.6 K. 
  }
\label{siocenter}
\end{figure}

\begin{figure}
\epsscale{0.9}
\plotone{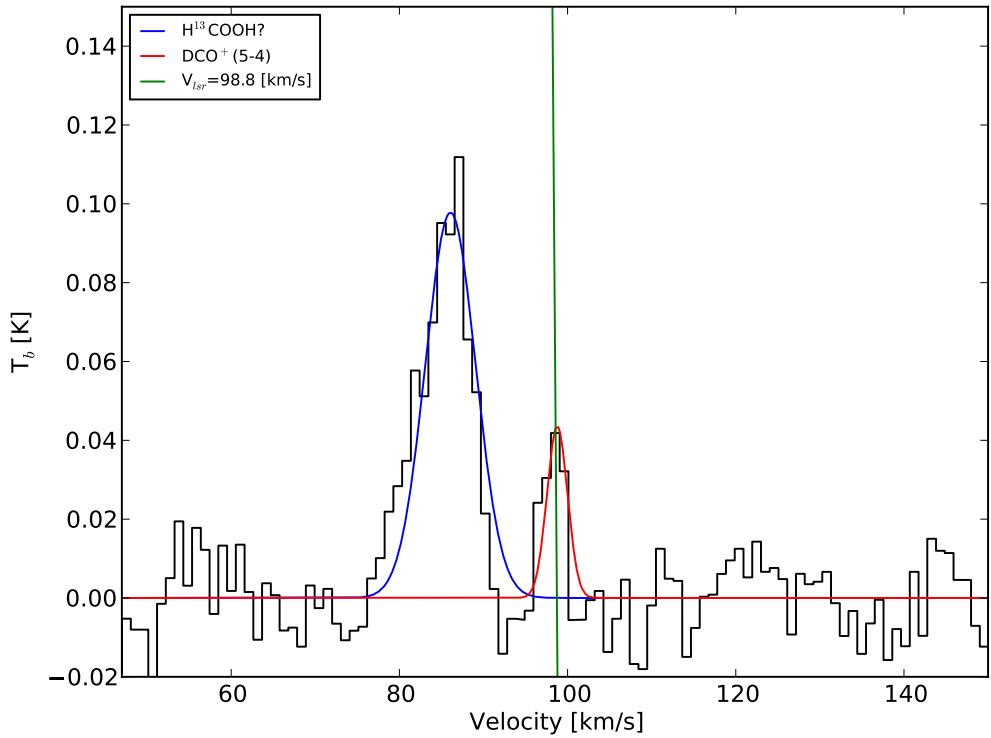}
\caption{
The spectrum shows our DCO$^{+}(5 \rightarrow 4)$ detection towards W43-MM1.
The Gaussian fit is shown by the red curve, note that the systemic
velocity of the clump, 98.9 \kms is shown by the green line. Also,
the un-identified line is also shown with its corresponding Gaussian fit
(blue curve). 
  }
\label{1}
\end{figure}

\begin{figure}
\centering
\includegraphics[width=0.9\hsize]{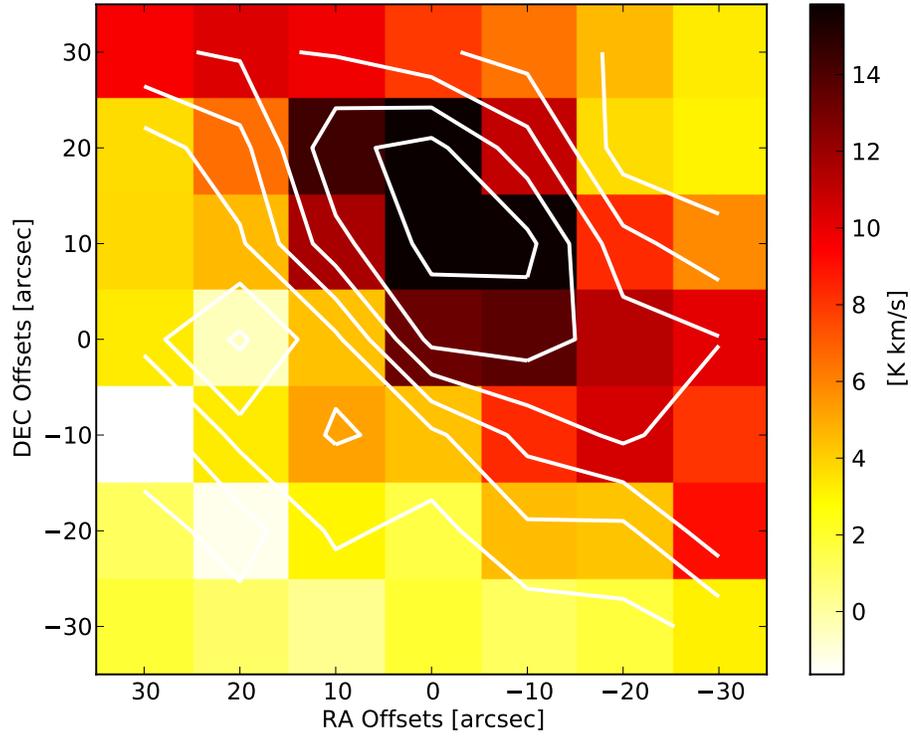}\caption{
A $60^{\prime \prime} \times 60^{\prime \prime}$ integrated intensity map
of N$_{2}$H$^{+}(4 \rightarrow 3)$ from W43-MM1 centered at the dust peak is shown. 
The color scale correspond to the integrated emission after applying a {\em nearest} interpolation
scheme in units of K \kms. The contours are chosen to be 2.5, 5, 7.5, 10, 12.5, and 15 K \kms. }
\label{n2hpmap}
\end{figure}

\begin{figure}
\centering
\includegraphics[width=0.9\hsize]{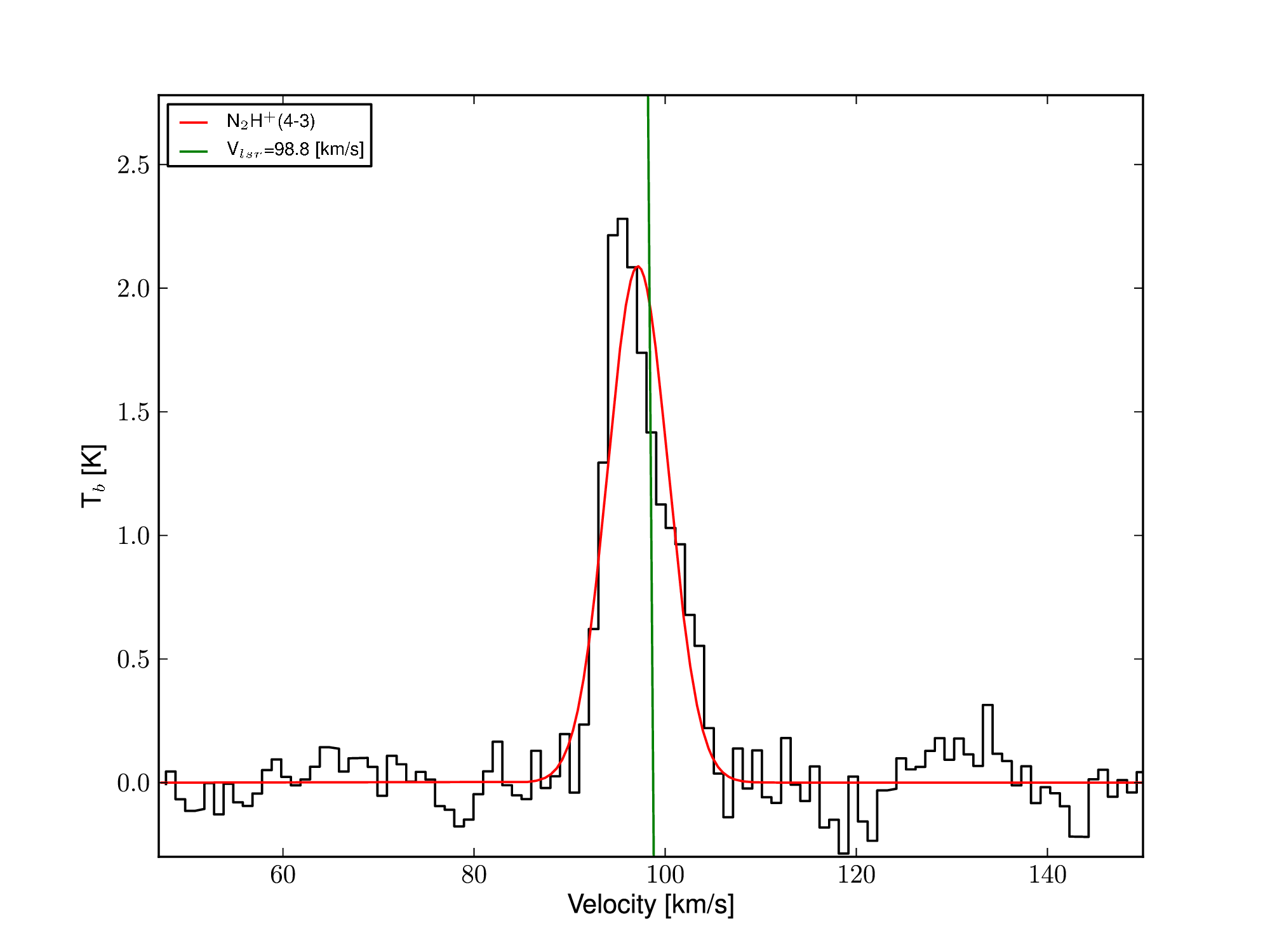}\caption{
The spectrum shows the N$_{2}$H$^{+}(4 \rightarrow 3)$ from the center 
of W43-MM1.
The Gaussian fit is shown by the red curve, note that the systemic
velocity of the clump, 98.9 \kms is shown by the green line. 
The spectra is binned every 10 channels giving a resolution in
velocity of about 1 \kms.
The spectrum was obtained 
with a higher signal to noise than the data presented in Figure \ref{n2hpmap} as we 
integrated additional time at the center of MM1. 
  }
\label{n2hpcenter}
\end{figure}

\begin{figure}
\centering
\includegraphics[width=0.9\hsize]{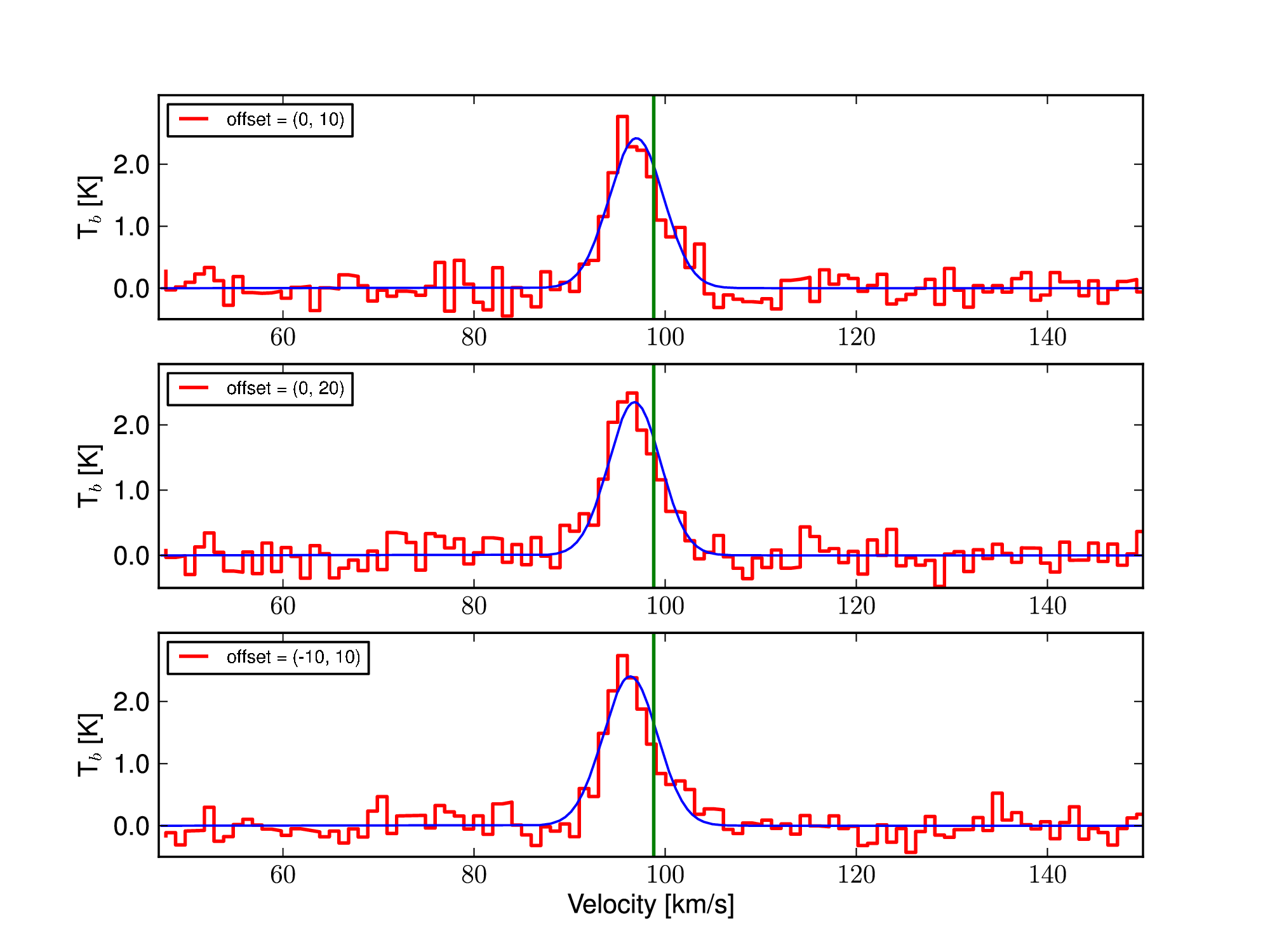}\caption{
 Spectra from N$_{2}$H$^{+}(4 \rightarrow 3)$
towards selected pointings in W43-MM1. The offsets are respect
to the main reference position an indicated by the label box in each
spectrum. The spectra is binned every 10 channels giving a resolution in
velocity of about 1 \kms.}
\label{n2hppanel}
\end{figure}

\begin{figure}
\epsscale{0.9}
\plotone{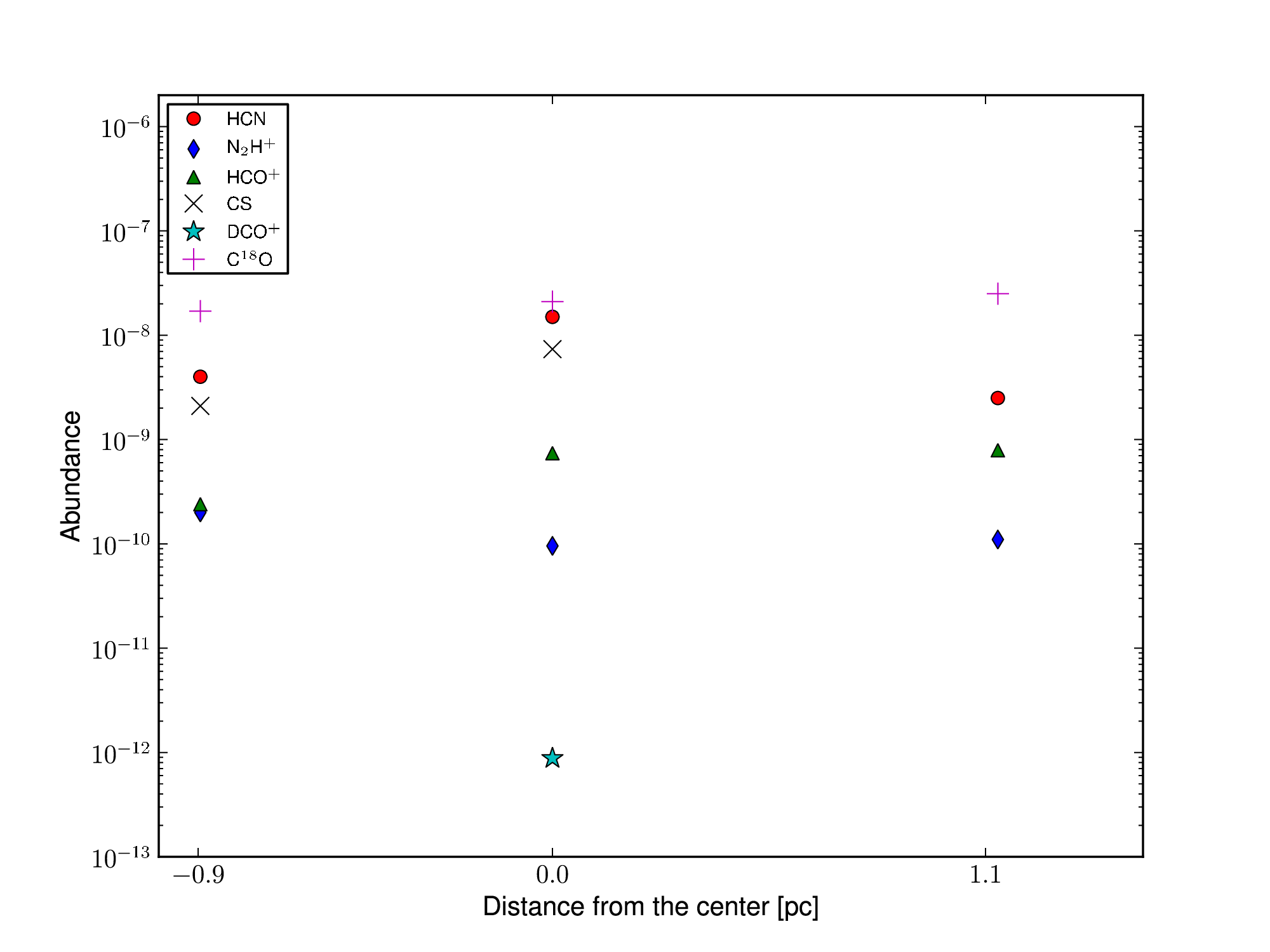}
\caption{The plot shows the fractional abundances derived for all the species observed  
at the three selected pointings along the MM1 strip (main axis). The points are indicated by their 
relative distance respect to the center. Thus, (30$^{\prime \prime}$, 30$^{\prime \prime}$) 
is at 1.1 pc, while (-30, -15) is at -0.9 pc from the  center. Each molecule is represented by
a symbol indicated in upper left box.
  }
\label{abundancePlot}
\end{figure}

To estimate abundances across the MM1 main axis,
we chose 3 pointings 
from north-east to south-west across MM1 main axis, including the center of MM1. 
The pointings are indicated as crosses in the
C$^{18}$O map (see Figure \ref{c18omap}).
Thus, we derived column densities along the filament for all available molecules
with the exception of DCO$^{+}$ which has only 1 data point and
CS which does not show emission over the $3\sigma$ level at the $(30^{\prime \prime}, 30^{\prime \prime})$
relative offset. 
The column density from the dust map obtained by \citet{Cortes2006a} was
used for the center pointing and 
new values were derived from the $350 \mu$m fluxes reported by \citet{Motte2003}
for  the remaining positions. 
The hydrogen column densities were derived using 

\begin{equation}
\label{cd}
(N_{\mathrm{H}}/\mathrm{cm}^{-2})=1.93 \times 10^{15}
\frac{(S_{\nu}/\mathrm{Jy})\lambda^{4}_{\mathrm{\mu m}}}
{(\theta_{s}/\mathrm{arcsec})^{2}(Z/Z_{\sun})bT}
\frac{e^{x} - 1}{x}
\end{equation}

\noindent where $S_{\nu}$ is the averaged densitydensity  flux from the source in Jy,
$\theta_{s}=
\sqrt{\theta_{s,min} \times \theta_{s,max}}$ is the
angular source size, ($Z/Z_\sun$=1) is relative metalicity to solar neighborhood,
$b$ is a parameter that reflects
the variation of dust absorption cross sections
\citep{Mezger1990}.
Two values are used for $b$; $b=1.9$ reproduces estimates of
$\sigma_{400\mu m} \sim 8.3
\times 10^{-26}$ cm$^{2}$(H-atom)$^{-1}$, which represents molecular gas
of moderate density $n({\textnormal H}_{2}) < 10^{6}$ cm$^{-3}$, and $b=3.4$
for dust around deeply embedded IR sources at higher densities. We used a value
of $b=3.4$ in our estimates for MM1. A value of
$T=19$ K is used for the dust temperature derived from the SED fitting by \citet{Motte2003},
  and  $x=\frac{1.44 \times 10^{4}}
{\lambda_{\mu m} T}$ is the $\frac{hc}{\lambda kT}$ factor for the Planck
function. 
The hydrogen column densities for the offsets are listed in Table \ref{pointings} 
and also indicated as crossed in the
C$^{18}$O map in Figure \ref{c18omap}.
The derived abundances for all molecules are listed in Table \ref{abundances}
and plotted in Figure \ref{abundancePlot}.

\subsection{Comparison with chemical models}

\subsubsection{HCN abundance}
The relative differences between molecular abundances in star forming regions can
provide information about the evolutionary state of that region. 
Chemical differentiation, like the N/O ratio, seems to be
ubiquitous in high mass star formation, possibly an indication of the evolutionary
state of a region \citep{Rodgers2001,Caselli2002b,Doty2002,Rodgers2003,Roueff2007}.
 Besides C$^{18}$O,
the most abundant molecule in our sample at the center of the MM1 clump appears to be HCN followed by 
CS, HCO$^{+}$, N$_{2}$H$^{+}$, and DCO$^{+}$. The difference between CS and
HCN is likely within the
calibration error, but larger for the remaining molecules. 
\citet{Jorgensen2004} studied a sample of low mass and high mass dense cores and found
the  HCO$^{+}$ abundance comparable to HCN, but lower than
the CS abundance in low-mass class I sources while in the high mass case HCN seems to be a factor of 5 
larger than HCO$^{+}$ but similar to the CS abundance. A similar situation is seen 
by \citet{Pirogov2007} in a study towards a sample of southern high mass star forming regions.
 Due to its high dipole moment, the
HCN molecule traces high density gas; its critical density for the $J=4\rightarrow 3$
transition is $n_{\mathrm{crit}} \sim 10^{7}$ cm$^{-3}$. Thus, it is expected to find
HCN$(4 \rightarrow 3)$ excited in dense clumps such as MM1. As the HCN molecule is most
likely formed in the gas phase rather than a product of ice mantle evaporation 
\citep{Rodgers2001}, its abundance might indicate how evolved the clump is.
It has been suggested that NH$_{3}$ would be a major component of ice mantles and
within the hot core region its release to the gas-phase will significantly affect 
the abundance of molecules such as HCN. As a daughter molecule,
HCN is formed by dissociative recombination of HCNH$^{+}$ which is previously formed 
by recombination of C$^{+}$ with NH$_{3}$. This scenario is plausible for higher gas
temperatures \citep{Rodgers2001} for models with $T \sim 300$ K 
in a hot core. Similar models have been produced by \citet{Doty2002} where they
find an enhanced HCN abundance as the hot core evolves with time, but using H$_{2}$NC$^{+}$
as the favored product to explain the HCN abundance. In principle the evaporation of
ammonia from ice mantles will reduce the abundance of O-bearing molecules and enhance the
abundance of N-bearing species.
Thus according to the previous studies, our derived abundance of HCN at the center 
give a possible age for the MM1 hot core of $10^{4}$ yr. 

\subsubsection{CS abundance}

The abundance of CS in evolved high mass clumps is likely dependent on the evaporation
of sulphur from ice mantles by radiation from the central star. As most of the sulphur
ejected from dust grains is likely locked into H$_{2}$S 
\citep{Doty2002,Rodgers2003}, or atomic ionized 
sulphur \citep{Wakelam2004}, this will lead to an increase of the SO$_{2}$ 
abundance, which will lead to  an increase in the CS abundance. Our derived value for the CS abundance towards the center
of MM1  is consistent with both models which place the CS
 abundance around $10^{-8}$ for an evolutionary
time scale of $10^{4}$ yr. However, the main issue behind using sulphur as a
 chemical clock is the uncertainty regarding the progenitors of S-bearing molecules which 
is still under  debate.

\subsubsection{Deuterium fractionation}

It is known that starless low mass cores have enhanced abundances of deuterium bearing molecules.
Due to freezing of carbon and oxygen on grain surfaces, the deuterium abundance can be enhanced 
by orders of magnitude with respect to more evolved regions \citep{Bergin2007}.  
Even though the situation in high mass clumps is not as clear,
it is expected that infrared dark clouds will show the same increase in 
deuterium abundance.
A useful parameter to quantify the importance of deuterium in clouds is the deuterium
fractionation.  This parameter is the ratio 
between the D-molecule and the H-molecule column densities and can be used to test the evolutionary
stage of star forming regions. Thus, we calculated R=[DCO$^{+}$]/[HCO$^{+}$] = 0.0012
for the center of MM1. Fractionation values have been calculated for low mass class 0 and class I
cores by \citet{Roberts2000,Turner2001,Caselli2002b} who found values larger than 0.01 with most
of the results about 0.06. Higher fractionation values based on deuterated formaldehyde,
about 0.1, have been obtained by \citet{Bergman2011} towards $\rho$-Ophiuchi.
In the case of high mass cores, results are not as numerous. \citet{Miettinen2009} found
moderate deuteration of about 0.03 to 0.04 for Orion B9; while \citet{Chen2010} found moderate
deuteration between 0.01 to 0.05 toward the infrared dark cloud G28.34+0.06.
\citet{Roueff2007} modeled deuterium fractionation in warm dense clumps, such as MM1, and found that
[DCO$^{+}$]/[HCO$^{+}$] decreases rapidly with temperature giving fractionation values similar to our findings for
T $< 30$ K, which suggests that MM1 might be more evolved than previously thought. 
Considering that infall is proceeding, as showed by PCI, and a hot core has already formed, 
it is possible that gas in the MM1 envelope has been heated by the hot core to the point 
where the dust and the gas are no longer in LTE. 
At the same time, it is also possible that the dust emission is optically thick.
In both cases the
dust temperature is not representative of the real kinetic gas temperature inside the 
MM1 clump.
Even though the amount of deuteration found in MM1 is not significant as
in pre-stellar sources, it is still significant for a warm environment such as
the one suggested by the model.
The DCO$^{+}$ molecule can be assembled by combination of CO with H$_{2}$D$^{+}$ which
is very efficient at low temperatures. 
However, chemical modeling by \citet{Roberts2000} suggests that
this mechanism is no longer efficient when temperature increases, which is also suggested by
\citet{Roueff2007}. As the envelope gas temperature rises the relative abundance of C, N, and O molecules
tend to decrease the abundance of H$_{2}$D$^{+}$ which can no longer efficiently combine
with CO to produce DCO$^{+}$. Instead, CH$_{2}$D$^{+}$ might become the dominant pathway for
DCO$^{+}$ creation by CH$_{2}$D$^{+}$ + O $\rightarrow$ DCO$^{+}$ + H$_{2}$
as the deuteration of CH$_{3}^{+}$ can proceed to much higher levels than H$_{2}$D$^{+}$, the usual way to produce 
DCO$^{+}$ (A. Wootten, priv. comm.). As previously mentioned, we question how well
the derived temperature from the SED represents the real envelope gas temperature; it
could well be warmer than the dust as the dust may have become optically thick.
So it is possible that 
the DCO$^{+}$ emission that we are seeing comes from a different chemical pathway than  
what is seen in colder sources.

\subsubsection{N$_{2}$H$^{+}$ and HCO$^{+}$ abundances}

For high mass star forming regions,
HCO$^{+}$ abundances have been derived from a sample of sources
similar to MM1 with values in the $10^{-11}$ to $10^{-10}$ range \citep{Zinchenko2009}.
H$^{13}$CO$^{+}$ column densities have been derived as well, toward several
high mass cores at different stages of evolution with values that
range between $10^{12}$ to $10^{13}$ cm$^{-2}$ \citep{Szymczak2007,Sakai2010}.
Our derived HCO$^{+}$ abundance towards the center of the MM1 clump is $X($HCO$^{+}$) = 
$1.5 \times 10^{-11}$ and the
H$^{13}$CO$^{+}$ column density is $2 \times 10^{13}$ which are consistent with
results in similar regions and with the chemical models previously described. 
The N$_{2}$H$^{+}$ abundance has also been derived towards
high mass cores, mostly from  N$_{2}$H$^{+}(1 \rightarrow 0)$ observations as there are few
N$_{2}$H$^{+}(4 \rightarrow 3)$ results in the literature. Current findings yield abundances
between $10^{-11}$ to $10^{-10}$ \citep{Daniel2007,Pirogov2007,Zinchenko2009,Miettinen2009}.
Our derived abundance towards the center of MM1 is $X($N$_{2}$H$^{+})$ = $9.6 \times 10^{-11}$ 
which is consistent with previous results. \citet{Busquet2011} produced a chemical model
for the high mass star forming region AFGL 5142. They modeled the
effect of a central source on the abundance of several species including 
N$_{2}$H$^{+}$. At an evolutionary time of $10^{4}$ yr, their model predicts an abundance
of $X($N$_{2}$H$^{+}) \sim 10^{-10}$ for the central core of AFGL 5142, which is about
the abundance value that we find for MM1 and in the same order of magnitude  with the
age obtained previously by comparing HCN, CS, and DCO$^{+}$ abundances with other chemical
models. \citet{Lintott2005} compared the production of
N$_{2}$H$^{+}$ and related chemical species for high mass star formation environments. While 
they do not present abundances for N$_{2}$H$^{+}$, they studied the relative abundance 
between CS and N$_{2}$H$^{+}$ which we can compare here. Their model predicts that
accelerated infall will produce a differentiation between CS, HCN, and N$_{2}$H$^{+}$, where
they expect that relative abundances will decrease due to the destruction of 
N$_{2}$H$^{+}$ by CO and enhancement of CS by the accelerated collapse.  
At the center of MM1 we obtained a log([CS]/[N$_{2}$H$^{+}$]) = 2.3, which in the \citet{Lintott2005} model
is close to no acceleration scenario. As MM1 is infalling with speed slightly supersonic
but sub-alfvenic (see section 6.6), it is difficult to reconcile this with a no
acceleration model.
However, we do see an enhancement of the CS abundance at the center of MM1
which will be discussed in the following section.

\begin{table*}
\centering                           
\caption[]{Column densities and abundances are presented here.} 
\label{abundances}      
\begin{tabular}{c c c c c}        
\hline\hline                 
Molecule & Offsets & Column Density & Abundance & Line-width\\     
         &  [(arcsec, arcsec)] &  [cm$^{-2}$] &  & \kms         \\
\hline                        
   DCO$^{+}$        &  (0,0) & $1.2 \times 10^{12}$ &  $8.8 \times 10^{-13}$ & 2.9\\   
   N$_{2}$H$^{+}$   &  (0,0) & $1.3 \times 10^{14}$ &  $9.6 \times 10^{-11}$ & 7.3\\
   N$_{2}$H$^{+}$   &  (30,30) & $6.0 \times 10^{13}$ &  $1.1 \times 10^{-10}$ & 6.3\\
   N$_{2}$H$^{+}$   &  (-30,-15) & $1.3 \times 10^{14}$ &  $2.0 \times 10^{-10}$ & 5.6\\
   HCO$^{+}$        &  (0,0) & $1.0 \times 10^{15}$ &  $7.4 \times 10^{-10}$ & 8.4   \\
   HCO$^{+}$        &  (30,30) & $2.5 \times 10^{14}$ &  $7.9 \times 10^{-10}$ & 6.2  \\
   HCO$^{+}$        &  (-30,-15) & $1.5 \times 10^{14}$ &  $2.4 \times 10^{-10}$ & 6.2   \\
   H$^{13}$CO$^{+}$ &  (0,0) & $2.0 \times 10^{13}$ &  $1.5 \times 10^{-11}$ & 3.0  \\
   HCN              &  (0,0) & $2.0 \times 10^{16}$ &  $1.5 \times 10^{-8}$ & 9.5\\
   HCN              &  (30,30) & $8.0 \times 10^{14}$ &  $2.5 \times 10^{-9}$ & 7.6\\
   HCN              &  (-30,-15) & $2.5 \times 10^{15}$ &  $4.0 \times 10^{-9}$ & 14 \\
   H$^{13}$CN       &  (0,0) & $4.0 \times 10^{14}$ &  $2.9 \times 10^{-10}$  & 4.8  \\
   CS               &  (0,0) & $1.0 \times 10^{16}$ &  $7.3 \times 10^{-9}$ &  4.3\\
   CS               &  (-30,-15) & $1.3 \times 10^{15}$ &  $2.1 \times 10^{-9}$ & 9.4\\
   C$^{18}$O        &  (0,0) & $2.9 \times 10^{16}$ &  $2.1 \times 10^{-8}$ & 6.4\\
   C$^{18}$O        &  (30,30) & $8.0 \times 10^{15}$ &  $2.5 \times 10^{-8}$ & 5.1 \\
   C$^{18}$O        &  (-30,-15) & $1.1 \times 10^{16}$ &  $1.7 \times 10^{-8}$ & 5.7\\
\hline                                    
\end{tabular}
\end{table*}

\begin{table*}
\centering                           
\caption[]{Hydrogen column densities for three selected pointings are shown here.
}
\label{pointings}      
\begin{tabular}{c c}        
\hline\hline                 
Offsets ($\alpha$, $\delta$) & Column Density  \\     
([arcsec], [arcsec]) &  [cm$^{-2}$]  \\
\hline                        
   (-30,-15) & $6.3 \times 10^{23}$ \\   
   (0,0)     & $1.5 \times 10^{24}$ \\
   (30,30)   & $3.2 \times 10^{23}$ \\
\hline                                    
\end{tabular}
\end{table*}

\subsection{Spatial differentiation}

In all the maps produced from our observations,
we see evidence for a filament in the MM1 clump (see also PCI and \citet{Motte2003}). 
The C$^{18}$O integrated emission map is consistent with this, as well as our abundance derivation which is 
constant along the main axis of the filament. 
The spatial distribution of both $^{12}$CO and dust (both 850 and 350 $\mu$m) also show
a single filament with a continuous progression in the emission from the MM1 center to the end of the 
filament at the south-east edge.
The $^{12}$CO map however, shows strong peaks in MM1 and
at the south-east edge where there is a likely boundary with the  \HII\  region. Only the MM1 center
peak is seen in dust emission maps. Our C$^{18}$O map shows
two distinct areas with strong emission around the dust peak and close to the south-east
boundary. There is a clear region in the middle where the emission drops in intensity, which is
not seen in the dust maps and it is not completely clear in the  $^{12}$CO map.
 This suggests that the MM1 filament might not extend
completely to the boundary with the \HII\ region as C$^{18}$O is likely showing us
the integrated emission through the cloud rather than just the envelope. The influence
of \HII\ region ionization and shock fronts as external pressure over the MM1 clumps
is important as a mechanism for  triggered star formation. 
The N$_{2}$H$^{+}$ emission is also distributed like C$^{18}$O and HCO$^{+}$, but we noticed that the peak emission in
N$_{2}$H$^{+}$ is not coincident with the dust peak as seen with the other molecules.
The main peak in the N$_{2}$H$^{+}$ integrated emission is 1 beam away from the center
to (0$^{\prime \prime}$, 20$^{\prime \prime}$). Thus, it cannot be attributed to an oversampled grid
re-distributing emission. \citet{Caselli2002b} suggested that N$_{2}$H$^{+}$ followed
the dust emission in cold cores as it has been seen that N-bearing species do not get
frozen on dust grains until densities reach the $10^{5}$ to $10^{6}$ range with 
gas temperatures around 10 K \citep{Bergin2007}. 
However, when the gas gets warmer due to heating
from the central source(s), CO is rapidly released from grain ice mantles quickly destroying
N$_{2}$H$^{+}$. If the gas in MM1 is being warmed by the hot core releasing CO into
the gas phase, it may explain the offset seen in the N$_{2}$H$^{+}$ peak emission with respect
to the dust peak. As our sub-millimeter observations trace  higher critical densities,
it is likely that we are seeing gas closer to the hot core where the temperatures
are likely higher than the dust SED suggested.

We also noticed that the emission from the neutral molecules
is more compact than the ions, with the exception of the H$^{13}$CO$^{+}$ and
H$^{13}$CN which only have significant emission from the center of MM1. 
The HCN and CS emission are mostly confined to the
clump, where CS is remarkably compact (see PCI Figures 7 and 8).
But, the N$_{2}$H$^{+}$ and HCO$^{+}$ integrated emission are much more widespread
over the filament. All these molecules have similar critical densities for the rotational
transitions observed; thus, they presumably come from gas under the same physical conditions. 
In order to explore possible explanations for this spatial differentiation, we will consider the
effect that the magnetic field threading MM1 might have over the gas along the filament. 
If the field is frozen in the ions, it is likely that the tension created by
the field will limit the movement of ions along the filament;
while the neutrals can diffuse more freely through the ions.
Table \ref{abundances} shows the line-widths obtained by adjusting Gaussian profiles to
the line emission at the selected positions along the strip. It is evident that the line-width
of N$_{2}$H$^{+}$ and HCO$^{+}$ are consistently smaller than
the neutral molecule line-widths. Particularly at the center we also have DCO$^{+}$ and
the $^{13}$C isotopomers of HCN and HCO$^{+}$. As these species are optically thin
their velocity dispersions are more representative than their more abundant $^{12}$C
isotopomers. All ions show smaller velocity dispersions than neutrals with
DCO$^{+}$ and H$^{13}$CO$^{+}$ about 3.0 \kms FWHM; while 
H$^{13}$CN has the largest line-width at 4.8 \kms. 
It has already been noted by \citet{Houde2000a,Houde2000b} that ions should have smaller velocity 
dispersion due to trapping on  magnetic field lines. \citet{Li2010} proposed that the smaller
velocity dispersion of ions with respect to neutrals, is 
a signature for turbulent ambipolar diffusion \citep{Li2010}. While ambipolar diffusion is
inevitable at some point during the clump evolution, 
how fast the neutrals will diffuse is still a matter of debate \citep{McKee2007}.

Chemical models of dense clumps, including inner hot cores, have calculated
radial profiles for the abundance of different molecules
\citep{Doty2002,Rodgers2003,Nomura2004}. All these models find that ion abundances increase 
rapidly with distance from the central source(s). Unfortunately, not all these models
present results for all the molecules presented here. Thus, we will concentrate on comparing with
the model that presents the largest number of molecular abundances. \citet{Rodgers2003} produced models
for envelopes surrounding hot cores, as we believe is the MM1 case. 
They developed models based on the dynamical state of the envelope, considering static and
collapsing cases  (unfortunately without including magnetic fields). 
For their collapse model and assuming gas temperatures of 300-400 K for the hot core,
they found HCO$^{+}$ abundances up to a few tenths of $10^{-9}$ at 1 pc from the central source(s)
for $10^{4}$ yr and $10^{-8}$ for $10^{5}$ yr of evolution after collapse started.
As PCI  presented compelling evidence of infall in MM1, comparing with this
model seems pertinent. Even though the UMIST database includes N$_{2}$H$^{+}$, they did not present 
predictions for that molecule in their work.
Table \ref{abundances} shows HCO$^{+}$ column densities and abundances at distances
of about 1 pc from the center at both sides of the filament. Our values are in the range 
of $10^{-10}$ to $10^{-11}$ which is at least an order of magnitude lower than the \citet{Rodgers2003}
model. The HCO$^{+}$ abundance is derived from H$^{13}$CO$^{+}$, which is moderately optically
thin (from the RADEX results, $\tau = 0.7$), but we assumed a $^{12}$CO/$^{13}$CO ratio of 50 which
is not necessarily accurate. However even using values close to 70 for the $^{12}$CO/$^{13}$CO ratio,
the HCO$^{+}$ abundance will not agree with the prediction.
Interestingly the
\citet{Rodgers2003} model predicts a sudden decrease in the abundance of neutral molecules such as CS, 
with a cutoff point around the collapse front. The case of HCN is not as clear, as their models
show a small decrease in abundance at the collapse front, but an increase towards the envelope.
The HCN abundance predicted at 1 pc from the center is about $5 \times 10^{-9}$,
 which is about what we see
from our data. For CS we derived an abundance similar to HCN at 1 pc, but the situation is
 not symmetric as we do not
have emission at the other end. This CS emission at offset $(-30^{\prime \prime}, -15^{\prime \prime})$
is likely coming from the same place where we see additional HCN emission ( see Figure 7 in PCI); 
thus, it would not be a valid comparison with the model. However, we do know that the CS emission in
MM1 is fairly compact which is also predicted by \citet{Rodgers2003}. 
The overall picture presented by the model is similar to what we have determined from our 
observations, particularly with the
strong CS cutoff at the collapse front.  Additionally, the panel presented by 
PCI for HCN infalling spectra places the collapse front at $25^{\prime \prime}$ in radius
(or about 0.9 pc ) which is also similar to the \citet{Rodgers2003} results. 
Even tough our abundances for neutral molecules agree with 
that model, the HCO$^{+}$ molecule does not.  However, it is feasible that 
changes in the chemical abundances due to the heating of the hot core explains the
spatial differentiation that we are seeing in the MM1 clump.

\subsection{The ionization rate}\label{se:io}

\subsubsection{The electron fraction}

The abundance of molecular ions depends on a complicated interplay between
gas-phase chemistry and cosmic-ray driven chemistry. Molecular ion abundances are 
critical to understand the interaction between the magnetic field and the
neutral gas as the magnetic field can only influence the dynamics of dense neutral gas
through the ions.
The electron fraction is estimated by charge equilibrium between the electrons and  the
most abundant ions in a molecular clump as we assume overall charge neutrality. 
The electron fraction permits calculations of neutral-ion coupling parameters, which we 
can use 
to understand the impact of the magnetic field in the gas dynamics of the infalling MM1 clump. 
As H$_{3}^{+}$, H$_{2}$D$^{+}$, DCO$^{+}$, HCO$^{+}$, N$_{2}$H$^{+}$, and N$_{2}$D$^{+}$
are considered the most abundant ions in a high mass clump
\citep{Miettinen2009,Caselli2008,Caselli2002b,Bergin1999,Williams1998,Wooten1979},
a lower limit for $\mathrm{X(e)}$ can be written as

\begin{equation}
\label{efrac}
\mathrm{X(e) \ge X(H_{3}^{+}) + X(H_{2}D^{+}) + X(DCO^{+}) + X(HCO^{+}) + X(N_{2}H^{+}) + X(N_{2}D^{+})}
\end{equation}

\noindent As previously shown, the deuterium fractionation is not large in MM1 as the clump
has already evolved from a pre-stellar phase. This suggests that deuterium species, such as
N$_{2}$D$^{+}$ and DCO$^{+}$, are not directly
important in the derivation of the electron fraction and can be neglected in the calculation.
The case of H$_{2}$D$^{+}$ is somewhat different, as
it is directly related to the H$_{3}^{+}$ abundance through 
$\mathrm{H_{3}^{+} + HD \leftrightarrow H_{2}D^{+} + H_{2}}$, and thus cannot be
neglected from equation \ref{efrac}.
The abundance of H$_{2}$D$^{+}$ can be estimated from the survey of \citet{Caselli2008} 
towards a sample
of star forming dense cores. In that work, the abundance of H$_{2}$D$^{+}$ was calculated
by observing the ortho-H$_{2}$D$^{+}(1_{1,0}-1_{1,1})$ line with CSO
\footnote{The CSO has a similar beam size as our ASTE observations}. 
To obtain an estimate of the H$_{2}$D$^{+}$ abundance, we used an average of their
values toward dense cores, or $X($H$_{2}$D$^{+}$) = $2.3 \times 10^{-10}$, and assumed
an ortho-to-para ratio of $10^{-4}$ as used by \citet{Miettinen2009}.
We are here assuming an average of the abundance values derived by  
\citet{Caselli2008} are representative for MM1.
Some dense cores used by them are indeed similar to MM1 (such as Ori B9), so using
an average is a compromise for a better representation of H$_{2}$D$^{+}$ abundance in
star forming regions. Now, the deuterium fraction $R$ provides a rough estimate of the
$r =$ [H$_{2}$D$^{+}$]/[H$_{3}^{+}$] abundance ratio through the relation derived by
\citet{Crapsi2004}, $R = (r + 2r^{2})/(3 + 2r + r^{2})$, 
by assuming a simplified chemical network for deuterium in molecular clouds. 
Thus,  by numerically solving the \citet{Crapsi2004}  relation and using 
$R = [\mathrm{DCO^{+}}]/[\mathrm{HCO^{+}}]$ = $1.2\times 10^{-3}$ from our observations, we obtain
$X($H$_{3}^{+}$) = $6.4 \times 10^{-8}$ giving an electron fraction of  $\mathrm{X(e)=6.5 \times 10^{-8}}$.
Our estimation for
the electron fraction is consistent with other high mass star forming sites
such as Ori 9 where \citet{Miettinen2009} calculated $6 \times 10^{-7}$ 
and DR21(OH) where \citet{Hezareh2008} obtained $3.2 \times 10^{-8}$, 
and in the upper range  of low mass
cores such as L1544 where \citet{Caselli2002b} found $\mathrm{X(e) \sim 10^{-9}}$.

\subsection{The Ion-Neutral coupling}

The effect of a  magnetic field in a weakly ionized gas depends on how well the
neutrals are coupled to the ions. A well coupled field-gas will be more efficient in
restraining the neutrals movement in directions parallel to the field
where magnetic support does not directly delay the gas infall. We first noticed that
infall in MM1 is slightly supersonic but sub-alfvenic, where 
$\mathrm{v_{if}} = 0.5$ \kms is the infalling speed, 
$\mathrm{C_{s}} = 0.46$ \kms is the sound speed, and $\mathrm{v_{A}} = 2.4$ \kms is
the Alfven speed. These parameters yield
a Mach number of M=1.1 and Alfvenic Mach number of $\mathrm{M_{A}} = 0.2$.
As the infall is sub-alfvenic, MHD waves can in principle propagate faster than the collapsing
gas. If the gas is warmer than the dust as previously suggested, the sound speed will be larger
making the infall sub-sonic. 

It has been found in laboratory plasmas that magnetic Reynolds numbers much larger
than 1 suggests that coupling between the gas and the field is strong and in principle, 
this allows the gas and the field to move as a single system. This number is defined
as the ratio between the inertial term to the viscous term in a magnetized fluid,
or $R_{m}=\frac{\mathrm{v_{if}}l}{\nu_{M}}$, where $\mathrm{v_{if}}$
is the fluid speed, $l$ is the length scale in which the magnetic field has a significant
change in magnitude, and $\nu_{M}$ is
the magnetic viscosity defined as $\nu_{M}=\mathrm{v_{A}}^{2}/\nu_{in}$, where 
$\nu_{in}$ is the ion-neutral collision frequency 
defined as $\nu_{in} = X(e)n_{\mathrm{H}_2}<\sigma_{in}v>$, where $<\sigma_{in}v>$ is the 
ion-neutral collision rate. Thus, we calculate the magnetic (ambipolar diffusion)
Reynolds number as $R_{m}=\frac{\mathrm{v_{if}}l}{\mathrm{v_{A}}^{2}}X(e)n_{H}<\sigma_{in}v>$, where 
$l$ is calculated to be about $\sim 5^{\prime \prime}$ or $ \sim 0.1$ pc obtained
 from the dust polarization map of
\citet{Cortes2006a}. The Alfven speed is  
$\mathrm{v_{A}}=2.2 \times 10^{5} B/(n_{H}A_{n})^{1/2}$ where
we take $A_{n}=1.4$ for representative molecular gas of cosmic abundance \citep{Zweibel2002}, 
$n_{H} = 4.5 \times 10^{5}$ cm$^{-3}$ from \citet{Cortes2006a}, and $B=855$ $\mu$G from
PCI. The $<\sigma_{in}v>$ is the
ion-neutral collision rate which is calculated using the Langeving approximation 
for the polarization potential \citep{McDaniel1973}. Here, we use a value of 
$<\sigma_{in}v>=1.69 \times 10^{-9}$ 
cm$^{3}$ s$^{-1}$ for molecular gas in a typical star forming region \citep{Mouschovias1981}.
Thus by using our derived values for $X(e)$, we calculate $R_{m} = 18.1$. 
As mentioned before, the ambipolar diffusion Reynolds number describes the stability of a magnetized flow, 
or how close the fluid is to become turbulent. The  boundary, or critical point, 
is often found for astrophysical fluid plasmas comparable to the Hartmann number
 $M$ \citep{Myers1995}, which
is defined as the ratio of the Lorentz force to the viscous forces
\citep{Davidson2001} and it can be written as $M=Bl\sqrt{\sigma / \eta}$, 
where $\sigma=c^{2}/(4\pi \nu_{M})$ is the
electrical conductivity and $\eta=\rho \nu$ is the dynamic viscosity function of
the kinematic viscosity $\nu$. \citet{Myers1995} derived the Hartmann number as
a function of column density as $M = 1.99 \times 10^{7}(N/10^{21} 
\mathrm{cm^{-2}}) X(e)^{1/2}$, the expression that we use here.
In astrophysical weakly ionized plasmas,  if $R_{m} \gg M \gg 1$ the flow and the field
can be considered to be turbulent as the dissipative term is not large enough to
damp the inertial term, but if $M > R_{m} \gg 1$ the field can still damp the
turbulence in the flow stopping the cascade into smaller eddies. 
Thus, we directly calculate the Hartmann number finding
$M=7 \times 10^{6}$. This value is consistent with calculations of \citet{Myers1995}
towards similar dense cores and to recent numerical simulations of sub-alfvenic flows 
\citep{LiMckee2008,Mckee2010}.

The ratio between the core size, $R$, and the
minimum MHD wavelength, or cutoff wavelenght, for propagation of MHD waves has
also been used to characterize the ion-neutral coupling. 
This ratio is known as the wave coupling number $W=R/\lambda_{0}$,
where  $\lambda_{0} = \frac{\pi V_{A}}{\nu_{in}}$, $V_{A}$  
is the Alfven speed, and $\nu_{in}$ is the ion-neutral 
collision frequency \citep{Elmegreen1993,Myers1995,Bergin1999}. Since  free MHD waves propagation
in a clump requires minimum wave damping, 
we expect this minimum wavelength to be less than the observed length-scale,
or $W > 1$, as this ratio gives the range between the maximum possible wavelength
(core size) and the cutoff wavelength \citep{Kulsrud1969,Elmegreen1993}. Thus,
when $W \gg 1$ we expect that the field and the gas are coupled over a large range of
length-scales.  Using our previous determination for the
strength of the magnetic field in this clump, 
the Alfven speed, and the ion-neutral collision frequency
we obtain a value for the wave coupling number of $W=110$.
Previous determinations of $W$ suggest that values close 1 are obtained
when the field is in equipartition with self-gravity; while greater values
are consistent with a dominant magnetic field and
strong coupling between the field and the neutral gas. Values around $W \sim 10^{3}$ have been 
found toward dense cores \citep{Myers1995}, but smaller values have also been 
obtained ($W \sim 20$) for massive cores \citep{Bergin1999}. Values smaller than $10^{3}$
suggest marginal coupling, where MHD waves can still propagate, but
damping by the field is still possible. We have obtained a result larger than 
the equipartition value, but still small in comparison with other dense cores.
As we have previously found, the mass-to-magnetic-flux ratio in this clump is super critical
and the dispersion function in the field lines, derived by PCI from the polarization map
of \citet{Cortes2006a}, suggests that a significant fraction of the
field is turbulent. Thus, a sligthly larger than 1 value for the ambipolar diffusion Reynolds number
and a larger than 1 wave coupling number value suggest that 
in MM1 the neutral gas has already or is in the process of  decoupling 
from the field and a transition
between a more dominant field and the full onset of gravitational collapse
is ongoing.  However, there is still sufficient dynamic range in the length-scales for the
gas and the field to move as a single system.
It is interesting that we are seeing an infalling clump with a field just decoupling
from the gas and a hot core already formed where molecular abundances are being
changed by the central source(s) and where no powerful outflows have been confirmed yet. 
This points to an intermediate scenario for the evolution of a high mass star forming region.
The clump seems to have evolved creating the inner source(s) before the envelope stops accreting and 
before the gas gets fully decoupled from the ambient magnetic field.
Moreover and even though MM1 is collapsing, we see evidence for an envelope in which the ions have 
a more extended distribution than the neutrals from species excited at similar critical densities. 
In principle ambipolar difussion can
distribute the magnetic flux to the envelope restricting the movement of ions while the neutrals
collapse. This mechanism can operate in the presence of MHD turbulence in which numerical
simulations suggests that the ambipolar diffusion time scale is significantly reduced, but
it can also efficiently redistribute magnetic flux \citep{LiMckee2008}. Nevertheless, we cannot
rule out that the spatial differentiation seen here is produced by chemistry alone, we favor
the previous mechanism as the more important one where chemical differentiation is
primarily responsible for the spatial differentiation seen in MM1. 

\section{Summary  and Conclusions}\label{se:Conclusion}

We have presented new and re-processed sub-millimeter data from W43-MM1.
New N$_{2}$H$^{+}(4 \rightarrow 3)$, DCO$^{+}(5 \rightarrow 4)$, and 
SiO$(8 \rightarrow 7)$  data have been obtained as well as previous un-published
C$^{18}$O$(3 \rightarrow 2)$ observations. We have re-processed
previous HCO$^{+}(4 \rightarrow 3)$, H$^{13}$CO$^{+}(4 \rightarrow 3)$,
HCN$(4 \rightarrow 3)$, H$^{13}$CN$(4 \rightarrow 3)$, and CS$(7 \rightarrow 6)$ data
to obtain chemical abundances in MM1.
New un-published results with the SMA, place a warm hot core, of about 400 K, inside MM1
as well as an outflow \citep{Sridharan2011}.
We did not detect SiO emission over the 3$\sigma$ threshold; thus, the outflow emission seen in
the SMA data is not seen from single dish measurements as it is likely deep inside the
clump or its emission is highly diluted within the single dish beam. 
We have calculated column densities by using 
the RADEX code, which agree with classical methods within a factor of 2.
Abundances for these lines have been derived for 3 selected points
 along the main axis of the MM1 clump. We have compared our findings with 
chemical models by \citet{Doty2002,Rodgers2003,Nomura2004,Roueff2007}.
Comparison with these models put the MM1 clump hot core
within an evolutionary age of about $10^{4}$ yr.
We also found the N$_{2}$H$^{+}(4 \rightarrow 3)$ peak emission offset from the dust peak by
 about 0.5 pc, which
can be explained by CO enhancement at the MM1 center due to the hot core as CO quickly destroys
N$_{2}$H$^{+}$. 
We derived a deuterium fractionation of 0.0012 based on our [DCO$^{+}$]/[HCO$^{+}$]
abundance ratio. This fractionation is consistent with an already evolved clump in
which the hot core is releasing C and O bearing molecules altering the chemical composition
of the MM1 clump and is also consistent with the depletion of N$_{2}$H$^{+}$ seen at the dust peak.
These findings question how well the dust temperature is coupled to the
gas temperature, as comparison with models suggest higher gas temperatures than the 19 K derived
from the dust SED. We conclude that the gas should be warmer than the dust inside the clump as the
dust emission might be likely optically thick.
At the same time, our maps of integrated emission suggest different spatial 
distributions for ions and neutral species; while the ions are more wide-spread around the
MM1 center, the neutrals (except C$^{18}$O) show a compact ditribution around the dust 
peak. It is possible that magnetic tension is restricting the ion movement while the
neutrals collapse more freely, but we favor chemical differentation induced by the hot core
as the main mechanism behind the spatial differentation seen in MM1. 
The C$^{18}$O map suggests that the MM1 clump does not
follow completely the filament suggested by the dust and the $^{12}$CO maps.
Two regions are seen, one corresponds to the MM1 clump and the other to what we believe
is the interface with the \HII\ region (a PDR?).
If this is confirmed, then the shock front from the \HII\ region has not yet reached
the MM1 clump, which may have implications for triggered star formation models.

By using our derived abundances for the ions, we have derived an 
electron fraction of $X(e) = 6.5 \times 10^{-8}$ where we used an average
value for the H$_{2}$D$^{+}$ abundance from \citet{Caselli2008} and our
deuterium fraction to derive the H$_{3}^{+}$ abundance. Our value for $X(e)$
is consistent with similar values in other
high mass star forming regions found in the literature. Using the PCI estimation for the
magnetic field strength, we calculated an ambipolar diffusion Reynolds number $R_{m} = 18.1$,
a Hartmann number $M= 7 \times 10^{6}$, and a wave coupling number $W=110$. These 
values suggest that the field has or is in the process of decoupling from the fluid as 
infall proceeds.
As the Reynolds number is still not comparable with the Hartmann number, the field
may still damp MHD waves, but as suggested by the wave coupling number, there is still enough
dynamic range in the length-scales for the field and the fluid to move as a single system.
By combining our findings here, we believe that the MM1 clump is in an intermediate
stage of evolution. At an intermediate stage of evolution, the inner source(s) has(have)
formed a hot core where a strong outflow is detected. The envelope is still contracting
and it is not fully decoupled from the ambient magnetic field.
To explore higher resolution
spatial scales, particularly at the cutoff (minimum wavelength) level, interferometric
observations will be needed. Thus, ALMA is the perfect instrument to study W43 further.

\begin{acknowledgements}
The National Radio Astronomy Observatory is a facility of the National Science Foundation operated under cooperative agreement by Associated Universities, Inc.
\end{acknowledgements}

\bibliographystyle{apj}
\bibliography{biblio}

\begin{thebibliography}{72}
\expandafter\ifx\csname natexlab\endcsname\relax\def\natexlab#1{#1}\fi

\bibitem[{{Arce} {et~al.}(2008){Arce}, {Santiago-Garc{\'{\i}}a},
  {J{\o}rgensen}, {Tafalla}, \& {Bachiller}}]{Arce2008}
{Arce}, H.~G., {Santiago-Garc{\'{\i}}a}, J., {J{\o}rgensen}, J.~K., {Tafalla},
  M., \& {Bachiller}, R. 2008, \apjl, 681, L21

\bibitem[{{Bergin} {et~al.}(1999){Bergin}, {Plume}, {Williams}, \&
  {Myers}}]{Bergin1999}
{Bergin}, E.~A., {Plume}, R., {Williams}, J.~P., \& {Myers}, P.~C. 1999, \apj,
  512, 724

\bibitem[{{Bergin} \& {Tafalla}(2007)}]{Bergin2007}
{Bergin}, E.~A. \& {Tafalla}, M. 2007, \araa, 45, 339

\bibitem[{{Bergman} {et~al.}(2011){Bergman}, {Parise}, {Liseau}, \&
  {Larsson}}]{Bergman2011}
{Bergman}, P., {Parise}, B., {Liseau}, R., \& {Larsson}, B. 2011, \aap, 527,
  A39+

\bibitem[{{Busquet} {et~al.}(2011){Busquet}, {Estalella}, {Zhang}, {Viti},
  {Palau}, {Ho}, \& {S{\'a}nchez-Monge}}]{Busquet2011}
{Busquet}, G., {Estalella}, R., {Zhang}, Q., {Viti}, S., {Palau}, A., {Ho},
  P.~T.~P., \& {S{\'a}nchez-Monge}, {\'A}. 2011, \aap, 525, A141+

\bibitem[{{Caselli} {et~al.}(2008){Caselli}, {Vastel}, {Ceccarelli}, {van der
  Tak}, {Crapsi}, \& {Bacmann}}]{Caselli2008}
{Caselli}, P., {Vastel}, C., {Ceccarelli}, C., {van der Tak}, F.~F.~S.,
  {Crapsi}, A., \& {Bacmann}, A. 2008, \aap, 492, 703

\bibitem[{{Caselli} {et~al.}(2002{\natexlab{a}}){Caselli}, {Walmsley},
  {Zucconi}, {Tafalla}, {Dore}, \& {Myers}}]{Caselli2002a}
{Caselli}, P., {Walmsley}, C.~M., {Zucconi}, A., {Tafalla}, M., {Dore}, L., \&
  {Myers}, P.~C. 2002{\natexlab{a}}, \apj, 565, 331

\bibitem[{{Caselli} {et~al.}(2002{\natexlab{b}}){Caselli}, {Walmsley},
  {Zucconi}, {Tafalla}, {Dore}, \& {Myers}}]{Caselli2002b}
---. 2002{\natexlab{b}}, \apj, 565, 344

\bibitem[{{Cesaroni} {et~al.}(1988){Cesaroni}, {Palagi}, {Felli}, {Catarzi},
  {Comoretto}, {di Francos}, {Giovanardi}, \& {Palla}}]{Cesaroni1988}
{Cesaroni}, R., {Palagi}, F., {Felli}, M., {Catarzi}, M., {Comoretto}, G., {di
  Francos}, {Giovanardi}, C., \& {Palla}, F. 1988, \aaps, 76, 445

\bibitem[{{Chen} {et~al.}(2010){Chen}, {Liu}, {Su}, \& {Zhang}}]{Chen2010}
{Chen}, H., {Liu}, S., {Su}, Y., \& {Zhang}, Q. 2010, \apjl, 713, L50

\bibitem[{{Cortes} {et~al.}(2008){Cortes}, {Crutcher}, \&
  {Bronfman}}]{Cortes2008}
{Cortes}, P., {Crutcher}, R.~M.~{Shepherd}, D., \& {Bronfman}, L. 2008, \apj,
  650, 1

\bibitem[{{Cortes} \& {Crutcher}(2006)}]{Cortes2006a}
{Cortes}, P. \& {Crutcher}, R.~M. 2006, \apj, 639, 965

\bibitem[{{Cortes} {et~al.}(2005){Cortes}, {Crutcher}, \&
  {Watson}}]{Cortes2005}
{Cortes}, P.~C., {Crutcher}, R.~M., \& {Watson}, W.~D. 2005, \apj, 628, 780

\bibitem[{{Cortes} {et~al.}(2010){Cortes}, {Parra}, {Cortes}, \&
  {Hardy}}]{Cortes2010}
{Cortes}, P.~C., {Parra}, R., {Cortes}, J.~R., \& {Hardy}, E. 2010, ArXiv
  e-prints

\bibitem[{{Crapsi} {et~al.}(2004){Crapsi}, {Caselli}, {Walmsley}, {Tafalla},
  {Lee}, {Bourke}, \& {Myers}}]{Crapsi2004}
{Crapsi}, A., {Caselli}, P., {Walmsley}, C.~M., {Tafalla}, M., {Lee}, C.~W.,
  {Bourke}, T.~L., \& {Myers}, P.~C. 2004, \aap, 420, 957

\bibitem[{{Crutcher} {et~al.}(1996){Crutcher}, {Troland}, {Lazareff}, \&
  {Kazes}}]{Crutcher1996}
{Crutcher}, R.~M., {Troland}, T.~H., {Lazareff}, B., \& {Kazes}, I. 1996, \apj,
  456, 217

\bibitem[{{Crutcher} {et~al.}(1999){Crutcher}, {Troland}, {Lazareff},
  {Paubert}, \& {Kaz{\` e}s}}]{Crutcher1999a}
{Crutcher}, R.~M., {Troland}, T.~H., {Lazareff}, B., {Paubert}, G., \& {Kaz{\`
  e}s}, I. 1999, \apjl, 514, L121

\bibitem[{{Daniel} {et~al.}(2007){Daniel}, {Cernicharo}, {Roueff}, {Gerin}, \&
  {Dubernet}}]{Daniel2007}
{Daniel}, F., {Cernicharo}, J., {Roueff}, E., {Gerin}, M., \& {Dubernet}, M.~L.
  2007, \apj, 667, 980

\bibitem[{{Davidson}(2001)}]{Davidson2001}
{Davidson}, P.~A. 2001, {An Introduction to Magnetohydrodynamics}, ed.
  {Davidson, P.~A.}

\bibitem[{{Doty} {et~al.}(2002){Doty}, {van Dishoeck}, {van der Tak}, \&
  {Boonman}}]{Doty2002}
{Doty}, S.~D., {van Dishoeck}, E.~F., {van der Tak}, F.~F.~S., \& {Boonman},
  A.~M.~S. 2002, \aap, 389, 446

\bibitem[{{Elmegreen} \& {Fiebig}(1993)}]{Elmegreen1993}
{Elmegreen}, B.~G. \& {Fiebig}, D. 1993, \aap, 270, 397

\bibitem[{{Fontani} {et~al.}(2006){Fontani}, {Caselli}, {Crapsi}, {Cesaroni},
  {Molinari}, {Testi}, \& {Brand}}]{Fontani2006}
{Fontani}, F., {Caselli}, P., {Crapsi}, A., {Cesaroni}, R., {Molinari}, S.,
  {Testi}, L., \& {Brand}, J. 2006, \aap, 460, 709

\bibitem[{{Friesen} {et~al.}(2010){Friesen}, {Di Francesco}, {Myers},
  {Belloche}, {Shirley}, {Bourke}, \& {Andr{\'e}}}]{Friesen2010}
{Friesen}, R.~K., {Di Francesco}, J., {Myers}, P.~C., {Belloche}, A.,
  {Shirley}, Y.~L., {Bourke}, T.~L., \& {Andr{\'e}}, P. 2010, \apj, 718, 666

\bibitem[{{Fuller} {et~al.}(2005){Fuller}, {Williams}, \&
  {Sridharan}}]{Fuller2005}
{Fuller}, G.~A., {Williams}, S.~J., \& {Sridharan}, T.~K. 2005, \aap, 442, 949

\bibitem[{{Garrod} {et~al.}(2008){Garrod}, {Weaver}, \& {Herbst}}]{Garrod2008}
{Garrod}, R.~T., {Weaver}, S.~L.~W., \& {Herbst}, E. 2008, \apj, 682, 283

\bibitem[{{G{\"u}sten} {et~al.}(2006){G{\"u}sten}, {Nyman}, {Schilke},
  {Menten}, {Cesarsky}, \& {Booth}}]{Gusten2006b}
{G{\"u}sten}, R., {Nyman}, L.~{\AA}., {Schilke}, P., {Menten}, K., {Cesarsky},
  C., \& {Booth}, R. 2006, \aap, 454, L13

\bibitem[{{Herbst} \& {van Dishoeck}(2009)}]{Herbst2009}
{Herbst}, E. \& {van Dishoeck}, E.~F. 2009, \araa, 47, 427

\bibitem[{{Hezareh} {et~al.}(2008){Hezareh}, {Houde}, {McCoey}, {Vastel}, \&
  {Peng}}]{Hezareh2008}
{Hezareh}, T., {Houde}, M., {McCoey}, C., {Vastel}, C., \& {Peng}, R. 2008,
  \apj, 684, 1221

\bibitem[{{Hoare} {et~al.}(2007){Hoare}, {Kurtz}, {Lizano}, {Keto}, \&
  {Hofner}}]{Hoare2007}
{Hoare}, M.~G., {Kurtz}, S.~E., {Lizano}, S., {Keto}, E., \& {Hofner}, P. 2007,
  in Protostars and Planets V, ed. B.~{Reipurth}, D.~{Jewitt}, \& K.~{Keil},
  181--196

\bibitem[{{Houde} {et~al.}(2000{\natexlab{a}}){Houde}, {Bastien}, {Peng},
  {Phillips}, \& {Yoshida}}]{Houde2000a}
{Houde}, M., {Bastien}, P., {Peng}, R., {Phillips}, T.~G., \& {Yoshida}, H.
  2000{\natexlab{a}}, \apj, 536, 857

\bibitem[{{Houde} {et~al.}(2000{\natexlab{b}}){Houde}, {Peng}, {Phillips},
  {Bastien}, \& {Yoshida}}]{Houde2000b}
{Houde}, M., {Peng}, R., {Phillips}, T.~G., {Bastien}, P., \& {Yoshida}, H.
  2000{\natexlab{b}}, \apj, 537, 245

\bibitem[{{Johnstone} {et~al.}(2010){Johnstone}, {Rosolowsky}, {Tafalla}, \&
  {Kirk}}]{Johnstone2010}
{Johnstone}, D., {Rosolowsky}, E., {Tafalla}, M., \& {Kirk}, H. 2010, \apj,
  711, 655

\bibitem[{{J{\o}rgensen} {et~al.}(2004){J{\o}rgensen}, {Sch{\"o}ier}, \& {van
  Dishoeck}}]{Jorgensen2004}
{J{\o}rgensen}, J.~K., {Sch{\"o}ier}, F.~L., \& {van Dishoeck}, E.~F. 2004,
  \aap, 416, 603

\bibitem[{{Kohno}(2005)}]{Kohno2005}
{Kohno}, K. 2005, in Astronomical Society of the Pacific Conference Series,
  Vol. 344, The Cool Universe: Observing Cosmic Dawn, ed. C.~{Lidman} \&
  D.~{Alloin}, 242--+

\bibitem[{{Kulsrud} \& {Pearce}(1969)}]{Kulsrud1969}
{Kulsrud}, R. \& {Pearce}, W.~P. 1969, \apj, 156, 445

\bibitem[{Lai(2001)}]{Lai2001}
Lai, S.~P. 2001, PhD thesis, University of Illinois at Urbana - Champaign,
  Urbana, IL 61801, available at the Astronomy library at the Astronomy
  building

\bibitem[{{Li} {et~al.}(2010){Li}, {Houde}, {Lai}, \& {Sridharan}}]{Li2010}
{Li}, H.-b., {Houde}, M., {Lai}, S.-p., \& {Sridharan}, T.~K. 2010, \apj, 718,
  905

\bibitem[{{Li} {et~al.}(2008){Li}, {McKee}, {Klein}, \& {Fisher}}]{LiMckee2008}
{Li}, P.~S., {McKee}, C.~F., {Klein}, R.~I., \& {Fisher}, R.~T. 2008, \apj,
  684, 380

\bibitem[{{Lintott} {et~al.}(2005){Lintott}, {Viti}, {Rawlings}, {Williams},
  {Hartquist}, {Caselli}, {Zinchenko}, \& {Myers}}]{Lintott2005}
{Lintott}, C.~J., {Viti}, S., {Rawlings}, J.~M.~C., {Williams}, D.~A.,
  {Hartquist}, T.~W., {Caselli}, P., {Zinchenko}, I., \& {Myers}, P. 2005,
  \apj, 620, 795

\bibitem[{{Liszt}(1995)}]{Liszt1995}
{Liszt}, H.~S. 1995, \aj, 109, 1204

\bibitem[{{Liu} {et~al.}(2002){Liu}, {Girart}, {Remijan}, \&
  {Snyder}}]{Liu2002}
{Liu}, S., {Girart}, J.~M., {Remijan}, A., \& {Snyder}, L.~E. 2002, \apj, 576,
  255

\bibitem[{{McDaniel} \& {Mason}(1973)}]{McDaniel1973}
{McDaniel}, E.~W. \& {Mason}, E.~A. 1973, {The Mobility and Diffusion of Ions
  in Gases} (New York Wiley-Interscience, 1973.)

\bibitem[{{McKee} {et~al.}(2010){McKee}, {Li}, \& {Klein}}]{Mckee2010}
{McKee}, C.~F., {Li}, P.~S., \& {Klein}, R.~I. 2010, \apj, 720, 1612

\bibitem[{{McKee} \& {Ostriker}(2007)}]{McKee2007}
{McKee}, C.~F. \& {Ostriker}, E.~C. 2007, \araa, 45, 565

\bibitem[{{Mezger} {et~al.}(1990){Mezger}, {Zylka}, \& {Wink}}]{Mezger1990}
{Mezger}, P.~G., {Zylka}, R., \& {Wink}, J.~E. 1990, \aap, 228, 95

\bibitem[{{Miettinen} {et~al.}(2009){Miettinen}, {Harju}, {Haikala},
  {Kainulainen}, \& {Johansson}}]{Miettinen2009}
{Miettinen}, O., {Harju}, J., {Haikala}, L.~K., {Kainulainen}, J., \&
  {Johansson}, L.~E.~B. 2009, \aap, 500, 845

\bibitem[{{Mooney} {et~al.}(1995){Mooney}, {Sievers}, {Mezger}, {Solomon},
  {Kreysa}, {Haslam}, \& {Lemke}}]{Mooney1995}
{Mooney}, T., {Sievers}, A., {Mezger}, P.~G., {Solomon}, P.~M., {Kreysa}, E.,
  {Haslam}, C.~G.~T., \& {Lemke}, R. 1995, \aap, 299, 869

\bibitem[{{Motte} {et~al.}(2003){Motte}, {Schilke}, \& {Lis}}]{Motte2003}
{Motte}, F., {Schilke}, P., \& {Lis}, D.~C. 2003, \apj, 582, 277

\bibitem[{{Mouschovias} \& {Paleologou}(1981)}]{Mouschovias1981}
{Mouschovias}, T.~C. \& {Paleologou}, E.~V. 1981, \apj, 246, 48

\bibitem[{Muller {et~al.}(2005)Muller, Schlˆder, Stutzki, \&
  Winnewisser}]{Muller2005}
Muller, H.~S., Schlˆder, F., Stutzki, J., \& Winnewisser, G. 2005, Journal of
  Molecular Structure, 742, 215 , mOLECULAR SPECTROSCOPY AND STRUCTURE - A
  Collection of Invited Papers in Honor of Dr. Walter J. Lafferty

\bibitem[{{Myers} \& {Khersonsky}(1995)}]{Myers1995}
{Myers}, P.~C. \& {Khersonsky}, V.~K. 1995, \apj, 442, 186

\bibitem[{{Nomura} \& {Millar}(2004)}]{Nomura2004}
{Nomura}, H. \& {Millar}, T.~J. 2004, \aap, 414, 409

\bibitem[{{Pirogov} {et~al.}(2007){Pirogov}, {Zinchenko}, {Caselli}, \&
  {Johansson}}]{Pirogov2007}
{Pirogov}, L., {Zinchenko}, I., {Caselli}, P., \& {Johansson}, L.~E.~B. 2007,
  \aap, 461, 523

\bibitem[{{Remijan} {et~al.}(2007){Remijan}, {Markwick-Kemper}, \& {ALMA
  Working Group on Spectral Line Frequencies}}]{Remijan2007a}
{Remijan}, A.~J., {Markwick-Kemper}, A., \& {ALMA Working Group on Spectral
  Line Frequencies}. 2007, in Bulletin of the American Astronomical Society,
  Vol.~38, Bulletin of the American Astronomical Society, 963--+

\bibitem[{{Roberts} \& {Millar}(2000)}]{Roberts2000}
{Roberts}, H. \& {Millar}, T.~J. 2000, \aap, 361, 388

\bibitem[{{Rodgers} \& {Charnley}(2001)}]{Rodgers2001}
{Rodgers}, S.~D. \& {Charnley}, S.~B. 2001, \apj, 546, 324

\bibitem[{{Rodgers} \& {Charnley}(2003)}]{Rodgers2003}
---. 2003, \apj, 585, 355

\bibitem[{{Roueff} {et~al.}(2007){Roueff}, {Parise}, \& {Herbst}}]{Roueff2007}
{Roueff}, E., {Parise}, B., \& {Herbst}, E. 2007, \aap, 464, 245

\bibitem[{{Sakai} {et~al.}(2010){Sakai}, {Sakai}, {Hirota}, \&
  {Yamamoto}}]{Sakai2010}
{Sakai}, T., {Sakai}, N., {Hirota}, T., \& {Yamamoto}, S. 2010, \apj, 714, 1658

\bibitem[{{Sch{\"o}ier} {et~al.}(2005){Sch{\"o}ier}, {van der Tak}, {van
  Dishoeck}, \& {Black}}]{Schoier2005}
{Sch{\"o}ier}, F.~L., {van der Tak}, F.~F.~S., {van Dishoeck}, E.~F., \&
  {Black}, J.~H. 2005, \aap, 432, 369

\bibitem[{{Smith} {et~al.}(1978){Smith}, {Biermann}, \& {Mezger}}]{Smith1978}
{Smith}, L.~F., {Biermann}, P., \& {Mezger}, P.~G. 1978, \aap, 66, 65

\bibitem[{{Sridharan to be submitted}(2011)}]{Sridharan2011}
{Sridharan to be submitted}, T.~K. 2011, \apj, 442, 186

\bibitem[{{Szymczak} {et~al.}(2007){Szymczak}, {Bartkiewicz}, \&
  {Richards}}]{Szymczak2007}
{Szymczak}, M., {Bartkiewicz}, A., \& {Richards}, A.~M.~S. 2007, \aap, 468, 617

\bibitem[{{Turner}(2001)}]{Turner2001}
{Turner}, B.~E. 2001, \apjs, 136, 579

\bibitem[{{Wakelam} {et~al.}(2004){Wakelam}, {Caselli}, {Ceccarelli}, {Herbst},
  \& {Castets}}]{Wakelam2004}
{Wakelam}, V., {Caselli}, P., {Ceccarelli}, C., {Herbst}, E., \& {Castets}, A.
  2004, \aap, 422, 159

\bibitem[{{Williams} {et~al.}(1998){Williams}, {Bergin}, {Caselli}, {Myers}, \&
  {Plume}}]{Williams1998}
{Williams}, J.~P., {Bergin}, E.~A., {Caselli}, P., {Myers}, P.~C., \& {Plume},
  R. 1998, \apj, 503, 689

\bibitem[{{Wilson} {et~al.}(1970){Wilson}, {Mezger}, {Gardner}, \&
  {Milne}}]{Wilson1970}
{Wilson}, T.~L., {Mezger}, P.~G., {Gardner}, F.~F., \& {Milne}, D.~K. 1970,
  \aap, 6, 364

\bibitem[{{Wilson} \& {Rood}(1994)}]{Wilson1994}
{Wilson}, T.~L. \& {Rood}, R. 1994, \araa, 32, 191

\bibitem[{{Wood} \& {Churchwell}(1989)}]{Wood1989}
{Wood}, D.~O.~S. \& {Churchwell}, E. 1989, \apjs, 69, 831

\bibitem[{{Wootten} {et~al.}(1979){Wootten}, {Snell}, \&
  {Glassgold}}]{Wooten1979}
{Wootten}, A., {Snell}, R., \& {Glassgold}, A.~E. 1979, \apj, 234, 876

\bibitem[{{Zinchenko} {et~al.}(2009){Zinchenko}, {Caselli}, \&
  {Pirogov}}]{Zinchenko2009}
{Zinchenko}, I., {Caselli}, P., \& {Pirogov}, L. 2009, \mnras, 395, 2234

\bibitem[{{Zweibel}(2002)}]{Zweibel2002}
{Zweibel}, E.~G. 2002, \apj, 567, 962

\end{thebibliography}

\end{document}